\documentclass[hyper,11pt,letterpaper]{JHEP3}

\usepackage{graphicx, amsmath,amssymb,calrsfs,mathrsfs} 

\newcommand{\mc}[1]{\mathscr{#1}}
\newcommand{\rv}[1]{\boldsymbol{\mathsf{#1}}}
\newcommand{\ra}{\rightarrow}

\newcommand{\vev}[1]{\left\langle #1\right\rangle}

\title{Universal Nonlinear Filtering Using Feynman Path Integrals II: The Continuous-Continuous Model with Additive Noise}

\author{Bhashyam Balaji\\
Radar Systems Section,\\
Defence Research and Development Canada, Ottawa,\\
3701 Carling Avenue, \\
Ottawa ON K1A 0Z4 Canada\\
Email: Bhashyam.Balaji@drdc-rddc.gc.ca\\
}

\maketitle

\abstract{
In this paper, the Feynman path integral formulation of the continuous-continuous filtering problem, a fundamental problem of applied science, is investigated for the case when the noise in the signal and measurement model is additive.  It is shown that it leads to an independent and self-contained analysis and solution of the problem. A consequence of this analysis is Feynman path integral formula for the conditional probability density that manifests the underlying physics of the problem. A corollary of the path integral formula is  the Yau algorithm that has been shown to be superior to all other known algorithms. The Feynman path integral formulation is shown to lead to practical and implementable algorithms. In particular, the solution of the Yau PDE is reduced to one of function computation and integration.}


\begin{document}

 \section{Introduction} 

 \subsection{Motivation}
The fundamental dynamical laws of physics, both classical and quantum mechanical, are described in terms of variables continuous in time. The continuous nature of the dynamical variables has been verified at all length scales probed so far, even though the relevant dynamical variables, and the  fundamental laws of physics, are very different in the microscopic and macroscopic realms. In practical situations, one often deals with macroscopic objects whose state variables  are phenomenologically well-described by classical deterministic laws modified by external disturbances that can be modelled as random noise. It is therefore natural to consider the problem of the evolution of a state of a signal of interest described by a continuous-time stochastic dynamical model. Roughly speaking, the state of the system is described by a noisy version of a deterministic dynamical system termed the state model, i.e., the dynamics is governed by a system of first-order differential equations in the state variable ($\rv{x}(t)$) with an additional contribution due to noise that is random($\rv{\nu}(t)$).  The noise in the state model is referred to as the signal noise. If the noise is Gaussian (or more generally  multiplicative Gaussian) the state process is a Markov process. Since the process is stochastic, the state process is completely characterized by a probability density function. The Fokker-Planck-Kolmogorov foward equation (FPKfe) describes the evolution of this probability density function (or equivalently, the transition probability density function) and is the complete solution of the state evolution problem. For an excellent discussion of stochastic processes and filtering theory, see \cite{A.H.Jazwinski1970}.  

However, in many applications the signal, or state variables, cannot  be directly observed. Instead, what is measured is a nonlinearly related stochastic process ($\rv{y}(t)$) called the measurement process. The measurement process can often be modelled as yet another continuous stochastic dynamical system called the measurement model. Loosely speaking, the observations, or measurements, are discrete-time samples drawn from a different system of noisy first order differential equations. The noise in the measurement dynamical system is referred to as measurement noise. For simplicity, the signal and measurement noise are often assumed to be independent. 

The problem of optimal nonlinear filtering is to estimate the state of a stochastic dynamical system optimal under a certain criterion, given the observations that are samples of a related stochastic measurement process. The conditional probability density function of the state parameters given the observations  is the complete solution of the filtering problem since it contains all the probabilistic information about the state process that is in the measurements and in the initial condition. This is the Bayesian approach, i.e., the  \textit{a priori} initial data  about the signal process contained in the initial probability distribution of the state($u(t_0,x)$) is incorporated into the solution. Given the conditional probability density, optimality may be defined under various criterion. Usually, the conditional mean, which is the least mean-squares estimate, is studied due to its richness in results and mathematical elegance. The solution of the optimal nonlinear filtering problem is termed universal, if the initial distribution can be arbitrary. 

\subsection{Fundamental Sochastic Filtering Results}

When the state and measurement processes are linear, the linear filtering problem was considered by Kalman and Bucy\cite{R.E.Kalman1960}, \cite{R.E.KalmanR.S.Bucy1961}. The Kalman filter has been successfully applied to a large number of problems.
 
Neverthless, the Kalman filter suffers from some major limitations. The Kalman filter is not optimal even for the linear filtering case if the initial distribution is not Gaussian\footnote{It may still be optimal for a linear system under certain criteria, such as minimum mean square error, but not a general criterion.}. In other words, the Kalman filter is not a universal optimal filter, even when the filtering problem is linear. In the general nonlinear case, the filter state is infinite dimensional, namely the whole conditional probability distribution function, and the Kalman filter cannot be optimal, although it may still work quite well if the nonlinearity is mild enough. 

The continuous-continous nonlinear filtering problem (i.e., continuous-time state and measurement stochastic processes) was studied in \cite{H.J.Kushner1964,H.J.Kushner1964a}, \cite{R.L.Stratanovich1960} and \cite{R.S.Bucy1965}. This led to a stochastic differential equation, called the Kushner equation, for the conditional probability density in the continuous-continuous filtering problem. It was noted in\cite{T.E.Duncan1967}, \cite{R.E.Mortensen1966}, and \cite{M.Zakai1969} that the stochastic differential equation satisfied by the unnormalized conditional probability density, called the Duncan-Mortensen-Zakai (DMZ) equation, is linear and hence considerably simpler than the Kushner equation. In \cite{M.H.A.Davis1980} and \cite{B.L.Rozovskii1972} were  derived the robust DMZ equation, a partial differential equation (PDE) that follows from the DMZ equation via a gauge transformation. 

A disadvantage of the robust DMZ equation is that the coefficients depend on the measurements. Thus, one does not know the PDE to solve prior to the measurements. As a result, real-time solution is impossible. In \cite{S.T.YauS.-TYau2000}, it was proved that the robust DMZ equation is equivalent to a partial differential equation that is independent of the measurements, which is referred to as the Yau Equation (YYe)  in this paper (see discussion at the end of Section \ref{sec:YauEquation} for why it is being distinguished from the FPKfe). Specifically, the measurements only enter as initial condition at each measurement step. Thus, no on-line solution of a PDE is needed; all PDE computations can be done off-line. As discussed later, there are other real-time solutions possible; however, the approximation error in those algorithms are not as well controlled. 

However, numerical solution of partial differential equations presents several challenges. A na\"ive discretization may not be convergent, i.e., the approximation error may not vanish as the grid size is reduced. Alternatively, when the discretization spacing is decreased, it may tend to a different equation, i.e., be inconsistent. Furthermore, the numerical method may be unstable. Finally, since the solution of the YYe is a probability density, it must be positive which may not be guaranteed by the discretization.  

A different approach to solving the PDE was taken in \cite{S.-T.YauS.S.-TYau1996} and \cite{S-T.YauS.S-T.Yau1998}. An explicit expression for the fundamental solution of the  YYe as an ordinary integral was derived. It was shown that the formal solution to the YYe may be written down as an ordinary, but somewhat complicated, multi-dimensional integral, with the integrand an infinite series. In addition, an estimate of the time needed for the solution to converge to the true solution was presented.  

\subsection{Outline of the Paper}

In this paper, the Feynman path integral (FPI) formulation is employed to tackle the continuous-continuous nonlinear filtering problem. Specifically, phrasing the stochastic filtering problem in a language common in physics, the solution of the stochastic filtering problem is presented. In particular, no other result in filtering theory (such as DMZ equation, robust DMZ equation, etc) is used. The path integral formulation leads to a path integral formula for the transition probability density for the general aditive noise case. A corollary of the FPI formalism is the path integral formula for the fundamental solution of the YYe and the Yau algorithm---the central result of nonlinear filtering theory. It is noted that this paper provides detailed derivation of results that were used in \cite{Balaji2008}. In view of the wide applicability of the results derived here, this paper provides a brief summary of the central results of filtering theory and pedagogical discussion of Feynman path itegral methods that may be found to be useful by a wider audience, i.e., people specializing in different areas of applied science and engineering.  

The Feynman path integral was introduced by Feynman in quantum mechanics\cite{R.P.FeynmanandA.R.Hibbs1965,Feynman1948}. It has had a long and impressive history of results in theoretical physics.  Inspired by an observation/remark of Dirac\cite{P.A.M.Dirac1933, P.A.M.Dirac1982}, Feynman discovered the path integral representation of quantum physics \cite{R.P.FeynmanandA.R.Hibbs1965,Feynman1948}. The importance of path integrals grew (in the mid 1960's) when Fadeev and Popov used the path integrals  to derive Feynman rules for the non-Abelian gauge theories. Its status was further enhanced when 't Hooft and Veltman used path integral methods to prove that the Yang-Mills theories were consistent (i.e., renormalizable). Since then, the path integral formulation of quantum field theory has led to extraordinary insights into quantum field theory, such as renormalization and renormalization group, anomalies, skyrmions, monopoles, instantons, and supersymmetry (see, for instance, the freely available text \cite{siegel-1999}). Much of the advance in the past 40 years in particle theory would not have been possible without the Feynman path integral. The path integral has also led to numerous insights into modern mathematics, especially in the work of Edward Witten.

A path integral formulation is especially attractive from a computational point of view. An excellent example is lattice quantum chromodynamics (QCD) (see, for instance, \cite{H.J.Rothe2005}). Specifically, lattice QCD has enabled high-precision calculations of observables in the non-Abelian gauge theory of strong interactions---an infinite dimensional, highly nonlinear problem. The numerical results derived using  the Feynman path integral based lattice QCD cannot be achieved by any other known technique. 

The following point needs to be emphasized to readers familiar with the discussion of standard filtering theory---Feynman path integral is different from the Feynman-K\v ac path integral. In filtering theory literature, it is the Feynman-K\v ac formalism that is used. The Feynman-K\v ac formulation is a rigorous formulation and has led to several rigorous results in filtering theory. However, in spite of considerable effort it has not been proven to be useful in the development of relaible practical algorithms with desirable numerical properties. It also obscures the physics of the problem.

In contrast, it is shown that the Feynman path integral leads to formulas that are eminently suitable for numerical implementation. It also provides a simple and clear physical picture. Finally, the theoretical insights provided by the Feynman path integral are highly valuable, as evidenced by numerous examples in modern theoretical physics (see, for instance, \cite{siegel-1999}), and shall be employed in subsequent papers. 

The outline of this paper is as follows. In Section \ref{sec:YauEqnReview}, aspects of continuous-continuous filtering are reviewed and  some of the recent work on continuous-continuous filtering is also discussed. In the following section, some topics in path integrals are reviewed and the interpretation of transition probability density as ensemble averages. For more details on the path integral methods, see any modern text on quantum field theory, such as  \cite{siegel-1999}, and especially \cite{Zinn-Justin2002} which discusses application of Feynman path integral to the study of stochastic processes. In Sections \ref{sec:PIFormula01} and \ref{sec:PIFormula02}, the path integral formulas for the transition probability density are derived for the time-independent and time-dependent cases. In Section \ref{sec:Verification}, the partial differential equation satisfied by the path integral formula (derived for sampled measurements) in the previous sections is derived. It is shown that in the time-dependent case the partial differential equation is simply the YYe of continuous-continuous filtering. The conclusions are presented in Section \ref{sec:Conclusion}. 

\section{Continuous-Continuous Filtering and the Yau Equation}\label{sec:YauEqnReview}
In this section, the main results of (continuous-continuous) nonlinear filtering theory are summarized. The most important results are the YYe and the Yau algorithm. In subsequent sections, it will be seen that they follow directly (and simply) from the Feynman path integral method.

 \subsection{The Continuous-Continuous Model}
The signal and observation model in continuous-continuous filtering is the following:
\begin{align}
  \label{eq:signalobse}
  {\left\lbrace\begin{aligned}  
  d\rv{x}(t)&=f(\rv{x}(t),t)dt+e(\rv{x}(t),t)d\rv{v}(t),\quad x(0)=x_0,\\ 
  d\rv{y}(t)&=h(\rv{x}(t),t)dt+n(t)d\rv{w}(t),\quad y(0)=0. \end{aligned}
\right.}
\end{align}
Here $\rv{x},\rv{v},\rv{y},$ and $\rv{w}$ are  $\mathbb{R}^n$, $\mathbb{R}^p$, $\mathbb{R}^m$ and $\mathbb{R}^q$ valued stochastic processes, respectively, and $e(\rv{x}(t),t)\in\mathbb{R}^{n\times p}$ and $n(t)\in\mathbb{R}^{m\times q}$. These are defined in the It\^o sense. The Brownian processes $\rv{v}$ and $\rv{w}$ are assumed to be independent\footnote{Hence, the large body of work on the correlated noise case is not considered in this paper} with $p\times p$ and $q\times q$ covariance matrices $Q(t)$ and $W(t)$, respectively. We denote $n(t)W(t)n^T(t)$ by $R(t)$, a $m\times m$ matrix. Also, $f$ is referred to as the drift, $e$ as the diffusion vielbein, and $eQe^T$ as the diffusion matrix. 

In this paper, the additive noise case is considered, i.e., $e(\rv{x}(t),t)=e(t)$. Then, the It\^o and Stratanovich forms are the same. In a future paper, the more general, multiplicative and correlated noise case shall be tackled. 

In this section, some of the relevant work on continuous-continuous filtering is summarized. Hence, it is assumed that $n=p, m=q$, $f$ and $h$ are vector-valued $C^{\infty}$ smooth functions, $e(\rv{x},t)$ is an orthogonal matrix-valued $C^{\infty}$ smooth function, $Q(t)$ is a $n\times n$ identity matrix, and $ n(t)$ and $R(t)$ are  $m\times m$ identity matrices. No explicit time dependence is assumed in the model.  

\subsection{The DMZ Stochastic Differential Equation} 
The unnormalized conditional probability density, $\sigma(t,x)$ of the state given the observations $\{Y(s):0\leq s\leq t\}$ satisfies the DMZ equation:
\begin{align}
  \label{eq:dmzsde}
  d\sigma(t,x)=\mc{L}_0\sigma(t,x)dt+\sum_{i=1}^m\mc{L}_i\sigma(t,x)dy_i(t),\quad \sigma(0,x)=\sigma_0.
\end{align}
Here 
\begin{align}
  \label{eq:L0}
  \mc{L}_0&=\frac{1}{2}\sum_{i=1}^n\frac{\partial^2}{\partial x_i^2}-\sum_{i=1}^nf_i(x)\frac{\partial}{\partial x_i}-\sum_{i=1}^n\frac{\partial f_i}{\partial x_i}(x)-\frac{1}{2}\sum_{i=1}^mh_i^2(x),\\ \nonumber
  &=-\sum_{i=1}^n\frac{\partial}{\partial x_i}\left( f_i(x)\cdot \right)+\frac{1}{2}\sum_{i=1}^n\frac{\partial^2}{\partial x_i^2}-\frac{1}{2}\sum_{i=1}^mh_i^2(x),\\ \nonumber
  &=\frac{1}{2}\left( \sum_{i=1}^nD_i^2-\eta(x) \right),
\end{align}
where $\mc{L}_i$ is the zero-degree differential operator of multiplication by $h_i(x), i=1,\ldots,m$, $\sigma_0$ is the probability density of the initial time $t_0$, and
\begin{align}
  \label{eq:defDeta}
  D_{i}&\equiv\frac{\partial}{\partial x_i}-f_i(x),\\ \nonumber
  \eta(x)&\equiv \sum_{i=1}^nf_i^2(x)+\sum_{i=1}^n\frac{\partial f_i}{\partial x_i}(x)+\sum_{i=1}^mh_i^2(x).
\end{align}
The DMZ equation is to be interpreted in the Stratanovich sense. Note that 
\begin{align}
	D_i&=\frac{\partial }{\partial x_i}-f_i(x),\\ \nonumber
	&=e^{F_i}\frac{\partial }{\partial x_i}e^{-F_i},\quad \text{where } F_i=F_i(x)=\int_0^xf_i(t)dt.
\end{align}
Hence, 
\begin{align}
	D_i^2=e^{F_i}\frac{\partial^2}{\partial x_i^2}e^{-F_i},
\end{align}
and
\begin{align}
	\mc{L}_0=\frac{1}{2}\left( \sum_{i=1}^ne^{F_i}\frac{\partial^2}{\partial x_i^2}e^{-F_i}-\eta(x) \right).
\end{align}

\subsection{The Robust DMZ Partial Differential Equation}

Following \cite{M.H.A.Davis1980}  and \cite{B.L.Rozovskii1972} introduce a new unnormalized density 
\begin{align}
  \label{eq:davistransf}
  u(t,x)=\exp\left( -\sum_{i=1}^mh_i(x)y_i(t) \right)\sigma(t,x).
\end{align}
Under this transformation, the DMZ SDE is transformed into the following time-varying PDE 
\begin{align}
\label{eq:pdedmz}
 {\left\lbrace\begin{aligned}
\frac{\partial u}{\partial t}(t,x)=&\frac{1}{2}\sum_{i=1}^n\frac{\partial^2 u}{\partial x_i^2}(t,x)
+\sum_{i=1}^n\left( -f_i(x)+\sum_{j=1}^my_j(t)\frac{\partial h_j}{\partial x_i}(x) \right)\frac{\partial u}{\partial x_i}(t,x)\\
&-\biggl( \sum_{i=1}^n\frac{\partial f_i}{\partial x_i}(x)+\frac{1}{2}\sum_{i=1}^mh_i^2(x)-\frac{1}{2}\sum_{i=1}^my_i(t)\Delta h_i(x)+\sum_{i=1}^m\sum_{j=1}^ny_i(t)f_j(x)\frac{\partial h_i}{\partial x_j}(x)\\ 
&\quad-\frac{1}{2}\sum_{i,j=1}^m\sum_{k=1}^ny_i(t)y_j(t)\frac{\partial h_i}{\partial x_k}(x)\frac{\partial h_j}{\partial x_k}(x) \biggr)u(t,x),\\  
u(0,x)=&\sigma_0(x).\end{aligned}
\right.}
\end{align}
This is called the robust DMZ equation. Here $\Delta$ is the Laplacian. The solution of a PDE when the initial condition is a delta function is called the fundamental solution.

\subsection{Universal Linear and Finite Dimensional Filtering I: Lie Algebra Method and the Maximal Rank Case}

The Kalman filter is not a universal optimal linear filter since it assumes that the initial distribution is Gaussian, although it may still be optimal under certain criteria such as minumum variance. In the general case, the DMZ equation needs to be considered. The use of Lie algebraic methods to solve the DMZ equation had been proposed in \cite{R.W.BrockettJ.M.C.Clark80} and \cite{R.W.Brockett81}. The Lie algebraic viewpoint has been useful in giving a deeper understanding of the DMZ equation. 

The estimation algebra $E$ associated with the continuous-continuous model is the Lie algebra generated by $\mc{L}_0,\mc{L}_1(x),\ldots,\mc{L}_m(x)$. It is said to be of maximal rank if for any $1\le i\le n$ there exist constants $c_i$ such that $x_i+c_i$ are in $E$. When $m\ge n$, if $h(x)=Hx$, where $H$ is of maximal rank, then $E$ is of maximal rank. Note that the explicit recursive solution for the Ben\v es filtering system (see \cite{Benes1981}) was originally derived only for the maximal rank case. In some cases, maximal rank condition is assumed in deriving the explicit solutions. 

If the linear filtering system is completely reachable, controllable, and observable, the estimation algebra is $2n+2$ dimensional with basis
\begin{align}
	\mc{L}_0,\frac{\partial}{\partial x_1},\ldots,\frac{\partial}{\partial x_n}, x_1,\dots, x_n,1.
\end{align}
The Wei-Norman approach to solve the robust DMZ equation, a time-variant PDE, is to note that the solution is of the form\cite{S.-T.Yau1994}
\begin{align}
	u(t,x)=e^{T(t)}e^{r_n(t)x_n}\cdots e^{r_1(t)x_1}e^{s_n(t)D_n}\cdots e^{s_1(t)D_1}e^{t\mc{L}_0}u(0,x),
\end{align}
where $T(t), r_1(t),\dots,r_n(t),s_1(t),\dots,s_n(t)$ satisfy a system of $2n+1$ ODEs. Therefore, the Wei-Norman approach requires the solution of a system of $(2n+1)$ ODEs, $n$ linear PDEs, and an FPKfe type of equation. However, if the linear system is not completely reachable or completely observable, the basis of the estimation algebra is not known beforehand and therefore needs to to calculated. Hence, one cannot even write down the finite system of ODEs explicitly.

\subsection{Universal Linear and Finite Dimensional Filtering II: Direct Solution}
A direct method for solving the universal linear filtering problem was first presented in \cite{YauYau1997}. It was shown that by a combination of a gauge transformation and a time-dependent spatial translation of the unnormalized conditional probability density, the robust DMZ equation for the linear filtering problem reduces to an on-line system of ODEs and an off-line FPKfe.  This solution does not require controllability or observability assumptions, and is also universal as the initial distribution could be arbitrary. This solution was further simplified and extended in \cite{G.Q.HuS.S-T.Yau2001} and \cite{G.Q.HuS.S.-T.Yau2002}. 

The direct solution for the Ben\v es case was also derived by noting that if $f$ is a vector field of a potential function $\phi$, then
\begin{align}
	\mc{L}_{0}=\frac{1}{2}\left( \sum_{i=1}^ne^{\phi}\frac{\partial^2}{\partial x_i^2}e^{-\phi}-\eta(x) \right).
\end{align}
If $\xi(t,x)=e^{-\phi}\sigma(t,x)$, then the DMZ equation is
\begin{align}
	\frac{\partial\xi}{\partial t}(t,x)&=\frac{1}{2}(\Delta-\eta(x))\xi(t,x)+\sum_{i=1}^m\mc{L}_i\xi(t,x)\frac{dy_i}{dt}(t),\\ \nonumber
	\xi(0,x)&=e^{-\phi}\sigma_0,
\end{align}
where $\Delta $ is the Laplacian operator. If $\eta$ is a quadratic polynomial in $x$, the semigroup generated by the differential operator $\Delta-\eta(x)$ is known and can be used to explicitly derive solutions to the equation when $h_i$ are linear in $x$. This technique is also related to the concept of equivalence of parabolic equations (see \cite{YauYau1997} for details). The solution was further generalized and simplified to the case of a more general finite-dimensional filter (Yau filter\footnote{A Yau filter is one where the drift in the signal model is the sum of a linear term and a gradient of a function $\phi(x)$ and the measurement model is linear. The function $\phi(x)$ is assumed to have no quadratic form piece; else it can be absorbed into the linear term. When the gradient term vanishes, the Kalman filter results. When the linear term vanishes and the gradient term is such that $\eta$ is quadratic, the Ben\v es filter\cite{Benes1981} is obtained. } with quadratic $\eta(x)$). In summary, they showed that if one considers $\tilde{u}(t,x)$ defined as
\begin{align}
\label{eq:fin-dimyau}
	\tilde{u}(t,x)=\exp\left[ c(t)+\sum_{i=1}^na_i(t)x_i-F(x+b(t)) \right]u(t,x+b(t)),
\end{align}
then the robust DMZ equation for $u(t)$ is reduced to a system of ODEs for $a(t), b(t),c(t)$, and a FPKfe-type of equation  for $\tilde{u}(t,x)$. It has since been shown that even the off-line FPKfe-type of equation  
can be reduced to a system of linear ODEs \cite{S.-T.YauLai2003,B.Balaji2006}. Note that, like the Lie algebra solution, the state and measurement models are  assumed to be  explicitly time-independent.

\subsection{The Yau Equation}\label{sec:YauEquation}

The finite dimensional filtering problem is a very restrictive class of the general nonlinear filtering problem. Recently, it was proved that the real-time solution of the general nonlinear filtering problem can be obtained reliably\cite{S.T.YauS.-TYau2000}, \cite{YauYau2008}. Let ${\cal{P}}=\left\{ \tau_0<\tau_1<\cdots<\tau_k=\tau \right\}$ be a partition of the time interval $[\tau_0,\tau]$, and let the norm of the partition ${\cal{P}}_k$ be defined as $|{\cal{P}}_k|=\sup_{1\le i\le k}\left\{ |\tau_i-\tau_{i-1}| \right\}$. If $u_l(t,x)$ satisfies the equation 
	\begin{align}
\label{eq:pdedmzfrozen}
 {\left\lbrace\begin{aligned}
\frac{\partial u_l}{\partial t}(t,x)=&\frac{1}{2}\sum_{i=1}^n\frac{\partial^2 u_l}{\partial x_i^2}(t,x)
+\sum_{i=1}^n\left( -f_i(x)+\sum_{j=1}^my_j(\tau_l)\frac{\partial h_j}{\partial x_i}(x) \right)\frac{\partial u_l}{\partial x_i}(t,x)\\
&-\biggl( \sum_{i=1}^n\frac{\partial f_i}{\partial x_i}(x)+\frac{1}{2}\sum_{i=1}^mh_i^2(x)-\frac{1}{2}\sum_{i=1}^my_i(\tau_l)\Delta h_i(x)+\sum_{i=1}^m\sum_{j=1}^ny_i(\tau_l)f_j(x)\frac{\partial h_i}{\partial x_j}(x)\\ 
&\quad-\frac{1}{2}\sum_{i,j=1}^m\sum_{k=1}^ny_i(\tau_l)y_j(\tau_l)\frac{\partial h_i}{\partial x_k}(x)\frac{\partial h_j}{\partial x_k}(x) \biggr)u_l(t,x),\\  
u_l(\tau_{l-1},x)=&u_{l-1}(\tau_{l-1},x),\end{aligned}
\right.}
\end{align}
in the time interval $\tau_{l-1}\le t\le\tau_l$, then the function $\tilde{u}_l(t,x)$ defined as
\begin{align}\label{eq:postveryau}
	\tilde{u}_l(t,x)=\exp\left( \sum_{i=1}^my_i(\tau_l)h_i(x) \right)u_l(t,x)
\end{align}
satisfies the parabolic partial differential equation
\begin{align}
	\label{eq:yauONFeqn}
	\frac{\partial \tilde{u}_l}{\partial t}(t,x)&=\frac{1}{2}\sum_{i=1}^n\frac{\partial^2\tilde{u}_l}{\partial x_i^2}(t,x)-\sum_{i=1}^nf_i(x)\frac{\partial \tilde{u}_l}{\partial x_i}(t,x)-\left( \sum_{i=1}^n\frac{\partial f_i}{\partial x_i}(x)+\frac{1}{2}\sum_{i=1}^mh_i^2(x) \right)\tilde{u}_l(t,x), \\ \nonumber
	&=-\sum_{i=1}^n\frac{\partial}{\partial x_i}\left[ f_i(x)\tilde{u}_l(t,x) \right]+\frac{1}{2}\sum_{i=1}^n\frac{\partial^2\tilde{u}_l}{\partial x_i^2}(t,x)-\frac{1}{2}\sum_{i=1}^mh_i^2(x)\tilde{u}_l(t,x),
\end{align}
in the same time interval. The converse of the statement is also true. In \cite{S.S.-T.YauandS.-T.Yau2001}, it was also noted that it is sufficient to use the previous observation, i.e., $u_l(t,x)$ satisfies Equation \ref{eq:pdedmzfrozen} if and only if $\tilde{u}_l(t,x)$ defined as
\begin{align}\label{eq:preveryau}
	\tilde{u}_l(t,x)=\exp\left( \sum_{i=1}^my_i(\tau_{l-1})h_i(x) \right)u_l(t,x)
\end{align}
satisfies Equation \ref{eq:yauONFeqn} in the time interval $\tau_{l-1}\le t\le\tau_l$.  We refer to Equations \ref{eq:postveryau} (\ref{eq:preveryau}) and Equation \ref{eq:yauONFeqn} as the post-measurement (pre-measurement) forms of the YYe.   

Observe that Equation \ref{eq:pdedmzfrozen} is obtained by setting $y(t)$ to $y(\tau_l)$ in Equation \ref{eq:pdedmz}\footnote{In contrast, note that $\tilde{u}(t,x)$ defined for solving the finite-dimensional  Yau filter in Equation \ref{eq:fin-dimyau} in the aforementioned papers is equivalent to solving the robust DMZ equation for $u(t,x)$ at \emph{each time instant}.}. It was proved that the solution of Equation \ref{eq:pdedmzfrozen} approximates very well the solution of the robust DMZ equation (Equation \ref{eq:pdedmz}), i.e., it converges to $u(t,x)$ in both pointwise sense and $L^2$ sense. Thus, solving Equation \ref{eq:pdedmz} is equivalent to solving Equation \ref{eq:yauONFeqn}.  Finally, 
\begin{align}
	\sigma(\tau,x)=\lim_{|{\cal{P}}_k|\rightarrow0}\tilde{u}_k(\tau_k,x).
\end{align}
Thus, the solution of the YYe (as $|{\cal{P}}_k|\rightarrow0$) is the desired unnormalized conditional probability density. 

Observe that when $h(x)=0$, it is simply the FPKfe. However, unlike the FPKfe, the YYe does not satisfy the current conservation condition, i.e., the right-hand term is not a total divergence. This means that it does not conserve probability. This fundamental difference is traced to the fact that \emph{the FPKfe evolves the normalized probability density (and preserves the normalization), while the YYe evolves the unnormalized conditional probability density}. Therefore, this distinction is made between the two equations in this paper.

 \subsection{The Yau Algorithm}

We may summarize the real-time algorithm, based on both the pre- and post-measurement forms of the YYe, of Yau as follows. Suppose measurements are available at times 
\begin{align}
	\cdots<\tau_0<\tau_1<\tau_2<\cdots<\tau_k=\tau.
\end{align}
We seek the solution $u_i(t,x)$, which is the solution of the robust DMZ equation. Let the initial distribution be $u(\tau_0,x)=\sigma_0(x)$. 
Then, evolve the initial distribution to the first measurement instant, $\tau_1$, using the YYe:
 \begin{align}
\label{eq:yauONFeqn00}
{\left\lbrace\begin{aligned}
\frac{\partial \tilde{u}_1}{\partial t}(t,x)&=\frac{1}{2}\sum_{i=1}^n\frac{\partial^2\tilde{u}_1}{\partial x_i^2}(t,x)-\sum_{i=1}^nf_i(x)\frac{\partial \tilde{u}_1}{\partial x_i}(t,x)-\left( \sum_{i=1}^n\frac{\partial f_i}{\partial x_i}(x)+\frac{1}{2}\sum_{i=1}^mh_i^2(x) \right)\tilde{u}_1(t,x),\\	
	\tilde{u}_1(\tau_0,x)=&
\begin{cases}
	\exp\left( \sum_{j=1}^m[y_j(\tau_0)-y_j(\tau_{-1})]h_j(x) \right)\sigma_0(x)&\text{(Pre-Measurement)}\\
	\exp\left( \sum_{j=1}^m[y_j(\tau_1)-y_j(\tau_0)]h_j(x) \right)\sigma_0(x)&\text{(Post-Measurement)}.
\end{cases}\end{aligned}
\right.}
\end{align}
The solution of equation \ref{eq:yauONFeqn00} at time $\tau_1$ is $\tilde{u}_1(\tau_1,x)$. Note that $u_1(\tau_1,x)$ is given by
\begin{align}
	u_1(\tau_1,x)=
\begin{cases}
	\exp\left( -\sum_{i=1}^my_i(\tau_0)h_i(x) \right)\tilde{u}_1(\tau_1,x)&\text{(Pre-Measuremnt)}\\
	\exp\left( -\sum_{i=1}^my_i(\tau_1)h_i(x) \right)\tilde{u}_1(\tau_1,x)&\text{(Post-Measurement)}.
\end{cases}
\end{align}
Next, solve the YYe to the next measurement instant $\tau_2$ with initial condition $\tilde{u}_2(\tau_1,x)$, i.e., 
\begin{align}
\label{eq:yauONFeqn01}
{\left\lbrace\begin{aligned}   
\frac{\partial \tilde{u}_2}{\partial t}(t,x)&=\frac{1}{2}\sum_{i=1}^n\frac{\partial^2\tilde{u}_2}{\partial x_i^2}(t,x)-\sum_{i=1}^nf_i(x)\frac{\partial \tilde{u}_2}{\partial x_i}(t,x)-\left( \sum_{i=1}^n\frac{\partial f_i}{\partial x_i}(x)+\frac{1}{2}\sum_{i=1}^mh_i^2(x) \right)\tilde{u}_2(t,x),\\	
\tilde{u}_2(\tau_1,x)&=
\begin{cases}
	\exp\left( \sum_{j=1}^m\left( y_j(\tau_1)-y_j(\tau_0) \right)h_j(x) \right)\tilde{u}_{1}(\tau_{1},x)&\text{(Pre-Measurement)}\\
	\exp\left( \sum_{j=1}^m\left( y_j(\tau_2)-y_j(\tau_1) \right)h_j(x) \right)\tilde{u}_{1}(\tau_{1},x)&\text{(Post-Measurement)}.
\end{cases}\end{aligned}
\right.}
\end{align}
to obtain $\tilde{u}_2(\tau_2,x)$. In fact, for $i\ge2$, $u_i(\tau_i,x)$ can be computed from $\tilde{u}_i(\tau_i,x)$, where $\tilde{u}_i(t,x)$ satisfies the equation
\begin{align}
\label{eq:yaustepi}
{\left\lbrace\begin{aligned}   	
\frac{\partial \tilde{u}_i}{\partial t}(t,x)&=\frac{1}{2}\sum_{i=1}^n\frac{\partial^2\tilde{u}}{\partial x_i^2}(t,x)-\sum_{j=1}^nf_j(x)\frac{\partial \tilde{u}}{\partial x_j}(t,x)-\left( \sum_{j=1}^n\frac{\partial f_j}{\partial x_j}(x)+\frac{1}{2}\sum_{j=1}^mh_{j}^2(x) \right)\tilde{u}_i(t,x),\\ 
	\tilde{u}_i(\tau_{i-1},x)&=
\begin{cases}
	\exp\left( \sum_{j=1}^m\left( y_j(\tau_{i-1})-y_j(\tau_{i-2}) \right)h_j(x) \right)\tilde{u}_{i-1}(\tau_{i-1},x)&\text{(Pre-Measurement)}\\
	\exp\left( \sum_{j=1}^m\left( y_j(\tau_{i})-y_j(\tau_{i-1}) \right)h_j(x) \right)\tilde{u}_{i-1}(\tau_{i-1},x)&\text{(Post-Measurement)}.
\end{cases}\end{aligned}
\right.}
\end{align}
The initial condition in Equation \ref{eq:yaustepi} follows from noting that (since $u_i(\tau_{i-1},x)=u_{i-1}(\tau_{i-1},x)$)
\begin{align}
	\tilde{u}_i(\tau_{i-1},x)&=u_i(\tau_{i-1},x)\exp\left( \sum_{j=1}^my_j(\tau_i)h_j(x) \right),\\ \nonumber
	&=\exp\left( -\sum_{j=1}^my_j(\tau_{i-1})h_j(x) \right)\tilde{u}_{i-1}(\tau_{i-1},x)\exp\left( \sum_{j=1}^my_j(\tau_i)h_j(x) \right),\quad\text{(Pre-Measurement)}\\
	\tilde{u}_i(\tau_{i-1},x)&=u_i(\tau_{i-1},x)\exp\left( \sum_{j=1}^my_j(\tau_i)h_j(x) \right),\\ \nonumber
	&=\exp\left( -\sum_{j=1}^my_j(\tau_{i-1})h_j(x) \right)\tilde{u}_{i-1}(\tau_{i-1},x)\exp\left( \sum_{j=1}^my_j(\tau_i)h_j(x) \right)\quad\text{(Post-Measurement)}.
\end{align}

Thus the solution of the robust DMZ equation at the $i$th step is given by
\begin{align}
	u_i(\tau_i,x)=
\begin{cases}
	\exp\left( -\sum_{j=1}^my_j(\tau_{i-1})h_j(x) \right)\tilde{u}_i(\tau_i,x)&\text{(Pre-Measurement)}\\
		\exp\left( -\sum_{j=1}^my_j(\tau_i)h_j(x) \right)\tilde{u}_i(\tau_i,x)&\text{(Post-Measurement)}.
\end{cases}
\end{align}

Note that the Yau algorithm is a recursive algorithm as it does not need any previous observation data. Furthermore, the YYe is independent of data and so can be computed off-line, and that the YYe is much simpler than the robust DMZ equation. Finally, note that the output of the Yau algorithm is the desired unnormalized conditional probability density. 

\subsection{Other Real-Time Algorithms}

 It is noted that there have been several other numerical methods proposed for solving the DMZ equation such as \cite{A.BensoussanR.GlowinskiA.Rascanu1990, R.J.ElliotR.Glowinski1989, P.FlorchingerF.LeGland1991,LototskyR.MikuleviciusB.L.Rozovskii1997}. Some numerical examples were presented in \cite{SunGlowinski1993}. As noted in \cite{SunGlowinski1993}, most papers on the approximate solution of the DMZ equation deal with algorithms that are not (or, at least not easily) implementable. The  most well-known of thsese ``implementable'' methods for the general filtering problem (rather than those applicable to only problems like finite-dimensional filters) are now briefly reviewed. 
 
Recall that the DMZ SDE for the unnormalized conditional probability density is given by 
\begin{align}\label{eq:DMZSDEItoform}
	d\sigma(t,x)=\mc{L}_I\sigma(t,x)+\sum_{i=1}^n\mc{L}_i\sigma(t,x)d\rv{y}_i(t),
\end{align}
where $\mc{L}_I=\mc{L}_0+\frac{1}{2}\sum_{i=1}^nh_i^2(x)$, i.e., $\mc{L}_I$ is the adjoint of the infinitesimal generator of the state Markov process, or the forward Fokker-Planck-Kolmogorov (FPK) operator. 

The simplest scheme is the explicit one-step Euler scheme:
\begin{align}
		\sigma(t+\Delta,x)=(1+\Delta\mc{L}_i)\sigma(t,x)+\sum_{i=1}^nh_i(x)(y_i(t+\Delta)-y_i(t))\sigma(t,x),
\end{align}
Some other algorithms follow from observing that Equation \ref{eq:DMZSDEItoform} can be considered as the combination of a linear stochastic equation and the FPKfe:
\begin{align}
	{\left\lbrace\begin{aligned}
		d\sigma_1(t,x)&=\sigma_1(t,x)\sum_{k=1}^mh_i(x)d\rv{y}_i(t),\\
		d\sigma_2(t,x)&=\mc{L}_I\sigma_2(t,x)dt.\end{aligned}
\right.}
\end{align}
The standard idea of operator splitting for solving deterministic higher dimensional PDEs can then be applied to this case. Since the linear stochastic equation can be solved explicitly, the following two splitting-up approximations result:
\begin{align}
	\sigma(t+\Delta,x)&=
{\left\lbrace\begin{aligned} 
		T_{\Delta}\exp\left( \sum_{i=1}^m[y_i(t+\Delta,x)-y_i(t,x)]h_i(x)-\Delta\frac{1}{2}\sum_{i=1}^mh_i^2(x) \right)\sigma(t,x),\\
		\exp\left( \sum_{i=1}^m[y(t+\Delta)-y(t)]h(x)-\Delta\frac{1}{2}\sum_{i=1}^mh_i^2(x) \right)T_{\Delta}\sigma(t,x).\end{aligned}
		\right.}
\end{align}
Here $T_{\Delta} $ is the operator corresponding to the evolution via the FPK forward equation (FPKfe). 

The spectral separation scheme  utilizes the fact that the conditional probability density can be written as a two-tensor basis with one of them obeying an observation-independent FPKfe type of equation (deterministic Hermite-Fourier coefficients) and the other a Wick polynomial that are completely defined by the observation process (see \cite{LototskyR.MikuleviciusB.L.Rozovskii1997} for details). This separation of parameters and observations also enables real-time implementation. It can be shown that it includes the explicit Euler one-step and splitting-up approximations.

Finally, it will be seen that the path integral formulas derived here can be used to solve the FPKfe arising in these techniques. In fact, this formula already arises in the investigation of the continuous-discrete filtering problem (see \cite{PAPER1}). Note that in the Yau algorithm, the ``prediction'' part is accomplished via the YYe. In contrast, in the Splitting-up method, it is done using the FPKfe, as in continuous-discrete filtering theory.  

\subsection{Why the Yau Algorithm is the most suitable algorithm}

The following two reasons explain why the Yau algorithm is superior to all other known algorithms: 

Firstly, recall that the coefficients of the robust DMZ equation (Equation \ref{eq:pdedmz}) contain the measurements. This means that the partial differential equation to solve for continuous-continuous filtering is not even known prior to the measurements. In other words, the robust DMZ equation has to be solved on-line; its solution cannot be pre-computed. In contrast, in the YYe (Equation \ref{eq:yauONFeqn}), measurements are absent from the partial differential equation. The measurements only enter the initial condition at each measurement step. This implies that the YYe can be be solved  off-line; there is no need for on-line solution of PDEs\footnote{This assumes that the measurement steps are equidistant or known.}. This makes real-time solution feasible even if state dimension is large. Specifically, suppose the measurements are performed at equidistant time intervals. The set of all possible initial conditions may be expressed as a linear combination of basis functions. Then, one may precompute and store the solution to the YYe for the basis functions. Upon arrival of measurements, the initial condition incorporating the measurements is computed. This is projected onto the basis functions to ascertain the components. The solution (due to linearity of the PDE), till the next measurement, is then simply the sum of the computed off-line solutions for the basis functions. 

Secondly, all other known algorithms assume the signal and measurement model drift were bounded. In fact, error bounds derived for these methods depended (often exponentially) on the bounds so that there are severe time-step restrictions. Note that these results cannot even cover the  Kalman filter (i.e., linear drift) case, which is a major limitation in solving practical problems. In contrast, the results of \cite{S.T.YauS.-TYau2000} and \cite{Shing-TungYauStephenS.T.Yau1998} assume only that the drifts are nonlinear with linear growth. The errors are independent of the bounds on the signal or measurement model drift. In \cite{YauYau2008}, this is further weakened: is is sufficient that the growth of the observation drift be greater than the growth of the signal model drift.

It will be seen that the Feynman path integral approach naturally leads to the Yau algorithm. In that sense, the Feynman path integral formulation gives a physical explanation why the Yau algorithm is the natural choice. In addition, the path integral formula explicitly solves the PDE.

\subsection{Integral Formula for the Solution of the YYe}

A solution of the YYe in terms of ordinary integrals was presented in \cite{S.-T.YauS.S.-TYau1996}, \cite{S-T.YauS.S-T.Yau1998}. The formal solution derived  is an infinite series in $t$ (for details see the reference \cite{S-T.YauS.S-T.Yau1998}). In addition, an estimate of the time interval on which this solution converges was presented. For the purposes of this paper, it is sufficient to note that it has better numerical properties (than solving PDEs) since integration can be computed reliably, as opposed to partial differential equations. However, the integrand is rather complicated (e.g., convolutions) and it does not appear to be easily implementable for a general model. We shall see that the path integral formula is considerably simpler to evaluate.

\section{Preliminaries}

In this section, a brief review of relevant aspects of functional methods common in theoretical physics is presented. For more details on these topics, the reader is referred to \cite{Zinn-Justin2002}.

\subsection{A Brief Review of Functional Calculus }

The infinite dimensional generalization of a function is called a functional, $F[x(t)]$, which gives a number for each function $x(t)$. 
Loosely speaking, the functional derivative may be defined analogous to the ordinary derivative as follows:
\begin{align}
	\frac{\delta F[x(t)]}{\delta x(t')}=\lim_{\epsilon\ra0}\frac{F[x(t)+\epsilon\delta(t-t')]-F[x(t)]}{\epsilon}.
\end{align}
Alternatively, the functional derivative, $\frac{\delta F}{\delta x(t')}[x(t)]$, of the functional $F[x(t)]$ with respect to variation of the function $x(t)$ at $s$ is defined  by the following equation
\begin{align} 
	F[x(t)+\eta(t)]=F[x(t)]+\int\frac{\delta F}{\delta x(s')}[x(s)]\eta(s')ds'+\cdots. 
\end{align}
Note that the functional derivative of a functional is also a functional. Functional differentiation obeys the standard algebraic rules (linearity and Leibnitz's rule):
\begin{align}
	\frac{\delta}{\delta x(t')}\left[ \sum_{i=1}^nF_{i}[x(t)] \right]&=\sum_{i=1}^n\frac{\delta F_i}{\delta x(t')}[x(t)],\\
	\frac{\delta}{\delta x(t')}\left[F_1[x(t)]F_2[x(t)]\right]&=F_1[x(t)]\frac{\delta}{\delta x(t')}F_2[x(t)]+F_2[x(t)]\frac{\delta}{\delta x(t')}F_1[x(t)].
\end{align}
Just as the derivative of $x$ with respect to $x$ is 1, the functional derivative of $x(t)$ with respect to $x(t')$ is the unit matrix in infinite dimensions, namely the delta function:
\begin{align}
	\frac{\delta x(t)}{\delta x(t')}=\delta(t-t').
\end{align}
The functional derivatives of a formal power series in function $x(t)$ can be seen to be analogous to the ordinary derivative of a power series in the variable $t$. Specifically, it is straightforward to verify that if 
\begin{align}
	F\left[ x(t) \right]= \sum_{n=0}^{\infty}\frac{1}{n!}\int dt_1\cdots dt_n F^{(n)}(t_1,t_2,\dots,t_n)x(t_1)\cdots x(t_n),
\end{align}
then 
\begin{align}
	F^{(r)}(t_1,t_2,\dots,t_r)=\left\{ \left( \prod_{i=1}^r\frac{\delta}{\delta x(t_i)} \right)F \right\}\Bigg|_{x=0}.
\end{align}
In particular, a functional representable as a power series in functions, like the exponential functional, can be functionally differentiated using this result. Thus
\begin{align}
	\frac{\delta}{\delta J(x)}\exp\left( \int dx'J(x')A(x') \right)=A(x)\exp\left( \int dx'J(x')A(x') \right).
\end{align}
The functional delta function may be written as
\begin{align}
	\label{eq:FuncDeltaFn}
	\delta(f(x(t))=\int\left[ d\lambda(t) \right]\exp\left(i\int dt\lambda(t)f(x(t))  \right).
\end{align}

In the application at hand, determinants of operators (matrices) arise in Gaussian integration in the infinite (finite) dimesional case. In particular, the operator $\det(\delta(x-y)+K(x,y))$ needs to be evaluated for the problem at hand, for some operator $K(x,y)$. It is assumed that traces of all powers of $K$ exist. From the identity
\begin{align}
	\text{ln}\det M=\text{tr}\text{ln}M,
\end{align}
it follows that
\begin{align}\label{eq:DetIdentity}
	\ln\det[1+K]=&\int dxK(x,x)-\frac{1}{2}\int dx_1dx_2K(x_1,x_2)K(x_2,x_3)+\cdots+\\ \nonumber
	&\frac{(-1)^{n+1}}{n}\int dx_1\cdots dx_n K(x_1,x_2)K(x_2,x_3)\cdots K(x_n,x_1)+\cdots.
\end{align}
Note that if $K(x,y)=\theta(x-y)\tilde{K}(x,y)$, only the first term is nonzero.

Finally, recall the the sequence of  identities from multivariable calculus
\begin{align}\label{eq:UnityFn}
	1&=\int \left\{ d^ny \right\}\delta^n(y),\\ \nonumber
	&=\int \left\{ d^nf(x) \right\}\delta^n(f(x)),\\ \nonumber
	&=\int\left\{  d^nx \right\} J\delta^n(f(x)),\quad J=\det\left( \frac{\partial f_i}{\partial x_j}(x) \right),\\ \nonumber
	&=\int \left\{  d^nx \right\}\delta^n(x-x_c).
\end{align}
Then, it follows that
\begin{align}
	\delta^n(f(x))&=\frac{\delta^n(x-x_c)}{J(x)},\\ \nonumber
	\delta^n(x-x_c)&=\delta^n(f(x))J(x).
\end{align}
and
\begin{align}
	\sigma(x)&=\int \left\{  d^nx \right\}\delta^n(f(x))J(x)\sigma(x).
\end{align}
Using the Fourier integral representation of the delta function, it is clear that 
\begin{align}
	\sigma(x)|_{f(x)=0}&=\frac{1}{(2\pi)^n}\int \left\{  d^nx \right\} \left\{  d^n\lambda \right\} e^{i\lambda f(x)}J(x)\sigma(x). 
\end{align}
The functional analogue of Equation \ref{eq:UnityFn} is
\begin{align}
\label{eq:UnityFunctional}
	1=\int[\mc{D}f(x(t')))\det\left[ \frac{\delta f}{\delta x(t')}(x(t))\right] \delta(f(x(t))).
\end{align}

Finally, the Gaussian integral identity
\begin{align}
	\int\left[ d^nx \right]\exp\left( -\frac{1}{2}\sum_{i,j=1}^nx_i(A^{-1})_{ij}x_j-\sum_{i=1}^nb_jx_j \right)=\exp\left( \frac{1}{2}\sum_{i,j=1}^nb_iA_{ij}b_j \right),\qquad \left[ d^nx \right]\equiv\frac{1}{\sqrt{(2\pi)^n\det A}},
\end{align}
becomes
\begin{align}
	\int[d\lambda(t)]\exp\left( -\frac{1}{2}\sum_{i,j=1}^n\int dt\lambda_i(t)\left( A^{-1}(t) \right)_{ij}\lambda_j(t)+\sum_{i=1}^n\int dtb_i(t)\lambda_i(t) \right)=\exp\left( \frac{1}{2}\sum_{i,j=1}^n\int b_i(t)A_{ij}(t)b_j(t) \right).
\end{align}

\subsection{Langevin Equation and Gaussian Measure}
In the path integral formulation, as in physics literature, the continuous-continuous model is interpreted as
\begin{align}
	\begin{cases}
		\dot{\rv{x}}(t)&=f(\rv{x}(t),t)+e(\rv{x}(t),t)\rv{\nu}(t),\\
		\dot{\rv{y}}(t)&=h(\rv{x}(t),t)+n(t)\rv{\mu}(t).
	\end{cases}
\end{align}
Such equations are called Langevin equations. Heuristically, 
\begin{align}
	\frac{d\rv{v}}{dt}\ra\rv{\nu}(t),\quad \frac{d\rv{w}}{dt}(t)\ra\rv{\mu}(t).
\end{align}

Gaussian noise process is represented by a path integral measure 
\begin{align}
\label{eq:GaussMeasure}
[d\rho(\nu(t))]=[\mc{D}\nu(t)]\exp\left[ -\frac{1}{2}\int dt\sum_{a,b=1}^p\nu_a(t)\left(Q^{-1}(t)\right)_{ab}\nu_b(t) \right], \quad \nu\in\mathbb{R}^{p\times1}.
\end{align}
where $\nu(t)$ is a real vector for each $t$. 

It is straightforward to verify that the mean is zero
\begin{align}
	\vev{\rv{\nu}_c(t)}&=\int[d\rho(\nu)]\nu_c(t),\\ \nonumber
	&=\int[{\mc{D}}\nu(t)]\nu_c(t)\exp{\left(-\frac{1}{2}\int dt\sum_{a,b=1}^p\nu_a(t)(Q^{-1}(t))_{ab}\nu_b(t)\right)},\\ \nonumber
	&=0,
\end{align}
and the variance is $Q(t)$\footnote{This is also consistent with the previous discussion in terms of It\^o calculus, i.e., $ d\nu_a(t)d\nu_b(t')=Q_{ab}(t)\delta_{tt'}dt$ implies that 
\begin{align}
 \vev{\nu_a(t)\nu_b(t')}\approx Q_{ab}(t)\frac{\delta_{tt'}}{\Delta t}\ra Q_{ab}(t)\delta(t-t').
\end{align}
}:
\begin{align}
	\vev{\rv{\nu}_c(t)\rv{\nu}_d(t')}&=\int[d\rho(\nu)]\nu_c(t)\nu_d(t'),\\ \nonumber
	&=\left[\int[\mc{D}\nu(t)]\nu_c(t)\nu_d(t')\exp{\left(-\frac{1}{2}\int dt\left[\sum_{a,b=1}^p\nu_a(t)(Q^{-1}(t))_{ab}\nu_b(t)+\sum_{a=1}^pb_a\nu_a(t)\right]\right)}\right]\Bigg|_{b_a=0},\\ \nonumber
	&=\frac{\delta}{\delta b_c(t)}\frac{\delta}{\delta b_d(t)}\exp\left(-\frac{1}{2}\int dt' \sum_{a,b=1}^pb_a(t')Q_{ab}(t')b_b(t')\right)\Bigg|_{b_a=0},\\ \nonumber
	&=Q_{cd}(t)\delta(t-t'),
\end{align}
and higher-order moments can be easily written down using Wick's theorem for bosonic (commuting) variables. The results are in accord with the expectation that the measure represents a Gaussian process with these first two moments. Finally, implicit in the use of these formal technques is the use of the Feynman convention, or symmetric discretization for the drift. 

\subsection{Transition Probability Density and Signal and Measurement Ensemble Averages}\label{sec:PrelimTrProbDen}

Consider an ensemble of systems with state variables described by the Langevin equation. Due to the random noise, each system leads to a different vector $x(t)$ that depends on time. Although only one realization of the stochastic process is ever observed, it is meaningful to speak about an ensemble average. For fixed times $t=t_i,i=1,2,\dots,r$, the probability density of finding the random vector $\rv{x}(t)$ in the ($n-$dimensional)interval $x_i\le \rv{x}(t_i)\le x_i+dx_i(1\le i\le r)$ is given by
\begin{align}
	W_r(t_r,x_r;\cdots;t_1,x_1)=\vev{\prod_{i=1}^r\delta^n(\rv{x}(t_i)-x_i)}, 
\end{align}
where $x_i$ is an $n-$dimensional column vector and $\vev{\cdot}$ denotes averaging with respect to the signal model noise $\rv{\nu}(t)$. The complete information on the random vector $\rv{x}(t)$ is contained in the infinite hierarchy of such probability densities. The quantity of interest here is the conditional probability density 
\begin{align}
	p(t_r,x_r|t_{r-1},x_{r-1},\ldots;t_1,x_1)&=\vev{\delta^{n}(\rv{x}(t_r)-x(t_r))}|_{x(t_{r-1})=x_{r-1},\dots,x(t_1)=x_1},\quad x(t_r)\equiv x_r\\ \nonumber
	&=\frac{W_r(t_r,x_r;\ldots;t_1,x_1)}{\int W_n(t_r,x_r;\ldots;t_1,x_1)d^nx_r}.
\end{align}

Now the process described by the Langevin equation with $\delta-$correlated Langevin force is a Markov process, i.e., the conditional probability density depends only on the value at the immediate previous  time:
\begin{align}
	p(t_r,x_r|t_{r-1},x_{r-1};\ldots;t_1,x_1)=p(t_n,x_n|t_{n-1},x_{n-1}).
\end{align}
Hence the complete information for a Markov process is contained in the transition probabilities
\begin{align}\label{eq:MarkovProp01}
	p(t_2,x_2|t_1,x_1)&=\frac{W_2(t_2,x_2;t_1,x_1)}{W_1(t_1,x_1)}, \\ \nonumber
	&=\vev{\delta^n(\rv{x}(t_2)-x_2)}|_{\rv{x}(t_1)=x_1}.
\end{align}
which satisfies the Chapman-Kolmogorov semigroup equation:
\begin{align}
	p(t_3,x_3|t_1,x_1)=\int p(t_3,x_3|t_2,x_2)p(t_2,x_2|t_1,x_1)dx_2. 
\end{align}

It can be shown that the transition probability satisfies the Fokker-Planck-Kolmogorov forward equation (FPKfe) (see for instance, \cite{H.Risken1999}) 
\begin{align}
	\frac{\partial p}{\partial t}(t,x)&=-\sum_{i=1}^n\frac{\partial }{\partial x_i}\left[f_i(x(t),t)p(t,x)\right]+\frac{1}{2}\sum_{i,k=1}^n\sum_{j=1}^p\frac{\partial^2 }{\partial x_i\partial x_j}\left[ (e(x(t),t)Q(t))_{ij}e_{jk}(x(t),t)p(t,x)) \right],\\ \nonumber
	&\equiv{\mc{L}}p(t,x).
\end{align}
In \cite{PAPER1}, the path integral formula for the fundamental solution for the FPkfe is derived and applied to the continuous-discrete filtering problem with additive (state model) noise.

In continuous-continuous filtering, the measurement stochastic process needs to be incorporated as well. Consider another ensemble of systems with state variables whose time evolution is governed by the measurement process. The measurement noise means that each system in the ensemble leads to a different time-dependent vector $y(t)$. Thus, even though only one realization of the measurement stochastic process is observed, it is still meaningful to talk about an ensemble average of the measurement process (in addition to one over the state process). Thus, the quantity of interest in continuous-continuous filtering is
\begin{align}
	P(t_2,x_2;y_2|t_1,x_1;y_1)=\vev{\vev{\delta^n(\rv{x}(t_2)-x_2)\delta^m(\rv{y}(t_2)-y_2)}}_{\rv{\mu}}|_{x(t_1)=x_1,y(t_1)=y_1},
\end{align}
where $\vev{\cdot}_{\rv{\mu}}$ denotes averaging with respect to the measurement noise $\rv{\mu}(t)$. In the following sections, the path integral formulas for $P(t_2,x_2;y_2|t_1,x_1;y_1)$ are derived. It shall be shown that the YYe plays the same role here that the FPKfe does in continuous-discrete filtering.

Note that the transition probability density is the complete solution to the continuous-continuous filtering problem, since if the initial distribution is $u(t_{i-1},x'|Y_{i-1})$, then the evolved conditional probability distribution is 
\begin{align}
	u(t,x|Y_i)=\int P(t,x;y_i|t_{i-1},x';y_{i-1})u(t_{i-1},x'|Y_{i-1})\left\{ d^nx' \right\}.
\end{align}
In Section \ref{sec:DerivYauAlg} it shall be shown that it leads to the Yau algorithm. 

\section{Path Integral Formula: Implicit Time Dependence}\label{sec:PIFormula01}

\subsection{General Result}
In this section, the following signal and measurement models are considered:
\begin{align}
	\begin{cases}
		d\rv{x}(t)&=f(\rv{x}(t))dt+d\rv{\nu}(t),\quad x(0)=x_0,\quad Q(t)=\hbar_{\nu}\mathbb{I}_{n\times n},\\
		d\rv{y}(t)&=h(\rv{x}(t))dt+d\rv{\mu}(t), \quad y(0)=0,\quad R(t)=\hbar_{\mu}\mathbb{I}_{m\times m}. 
\end{cases}
\end{align}
Here $\hbar_{\nu}$ and $\hbar_{\mu} $ are positive real numbers and $\mathbb{I}_{n\times n}$ is the $n\times n$ identity matrix. This is the same as the model in \cite{S.T.YauS.-TYau2000}, except in those references $\hbar_{\mu}=\hbar_{\nu}=1$, and an orthogonal vielbein matrix multiplied the state Brownian process. 

Consider the member of the state ensemble with initial state $x(t_0)=x_0$ and final state $x(t)=x$ and with signal noise sample history $\nu_i(t), i=1,\dots,n$. Then, from the state equation 
\begin{align}
	\dot{x}_i(t)=f_i(x(t))+\nu_i(t),\quad x(t_0)=x_0, x(t)=x.
\end{align}
Similarly, consider a member of the measurement ensemble with measurement noise sample history $\mu_i(t),i=1,\dots,m$. Then,  the measurements satisfy the equation 
\begin{align}
	\dot{y}_k(t)=h_k(x(t))+\mu_k(t),\quad y(t_0)=y_0,y(t)=y, 
\end{align}
and $k=1,\dots,m$. The transition probability density is computed by averaging over the signal and measurement ensembles and imposing the condition that the signal and measurement Langevin equations are satisfied. That is,
\begin{align}
	P(t,x;y|t_0,x_0;y_0)&=\int[d\rho(\nu(t))][d\rho(\mu(t))]\\ \nonumber
	&\qquad\times [d\dot{x}(t)]J\delta^n[\dot{x}(t)-f(x(t))-\nu(t)]\\ \nonumber
	&\qquad\times [d\dot{y}(t)] J_y\delta^m[\dot{y}(t)-h(x(t))-\mu(t)]\\ \nonumber
	&\qquad\times\delta^n[x(t)-x]|_{x(t_0)=x_0}\delta^m[y(t)-y]|_{y(t_0)=y_0}.
\end{align}
Here $J$ and $J_y$ are Jacobians that are computed below. Note that the measurement Langevin equation is imposed in the path integral expression. 

Based on the assumptions of the signal and measurement noise processes, it is clear that 
\begin{align}
	[d\rho(\nu(t))]&=[\mc{D}\nu(t)]\exp\left( -\frac{1}{2\hbar_{\nu}}\sum_{i=1}^n\int_{t_0}^t\nu_i^2(t)dt \right), \\ 
	[d\rho(\mu(t))]&=[\mc{D}\mu(t)]\exp\left( -\frac{1}{2\hbar_{\mu}}\sum_{i=1}^n\int_{t_0}^t\mu_i^2(t)dt \right).
\end{align}

Next, the Jacobians, $J$ and $J_y$ are computed. Observe that  
\begin{align}
\label{eq:Jacobian00}
	\frac{\delta}{\delta x_j(t')}\left[ \dot{x}_i(t)-f_i(x(t))-\nu_i(t) \right]&=\left[ \delta_{ij}\frac{d}{dt}- \frac{\partial f_i}{\partial x_j}(x(t))\right]\delta(t-t'),\\ \nonumber
	&=-\frac{d}{dt'}\left[ \delta_{ij}\delta(t-t')-\theta(t-t')\frac{\partial f_i}{\partial x_j}(x(t)) \right],  \nonumber
\end{align}
since
\begin{align}
	\frac{d}{dt'}\theta(t-t')=-\delta(t-t'),\quad\text{ and }\quad \frac{\delta x(t)}{\delta x(t')}=\delta(t-t').
\end{align}
The Jacobian $J$ is thus given by 
\begin{align}
	\label{eq:Jacobian01}
	J&\equiv\det\left( -\frac{d}{dt'}\left[ \delta_{ij}\delta(t-t')-\theta(t-t')\frac{\partial f_i}{\partial x_j}(x(t)) \right] \right),\\ \nonumber
	&=\det\left( -\frac{d}{dt'} \right)\det\left( \delta_{ij}\delta(t-t')-\theta(t-t')\frac{\partial f_i}{\partial x_j}(x(t)) \right),\\ \nonumber
	&=\mc{N}\det\left( \delta_{ij}\delta(t-t')-\theta(t-t')\frac{\partial f_i}{\partial x_j}(x(t) )\right),  
\end{align}
where $\mc{N}$ is an irrelevant constant (independent of $x$) that may be absorbed in the measure. From the identity $\det A=\exp\left( \text{tr}\ln A \right)$, and using the identity in Equation \ref{eq:DetIdentity} 
\begin{align}
	\label{eq:Jacobian02}
	\text{ln}J&=\ln\det\left[ \delta_{ij}\delta(t-t')-\theta(t-t')\frac{\partial f_i}{\partial x_j}(x(t) \right],\\ \nonumber
	&=-\theta(0)\int dt\sum_{i=1}^n\frac{\partial f_i}{\partial x_i}(x(t),\\ \nonumber
	&=-\frac{1}{2}\int\sum_{i=1}^n\frac{\partial f_i}{\partial x_i}(x(t))dt.
\end{align}
The choice $\theta(0)=\frac{1}{2}$ ensures that commutativity of averaging and time differentiation \cite{Zinn-Justin2002}. Similarly, it is straightforward to see that the Jacobian $J_y$ is trivial since $\dot{y}(t)$ is $y-$independent, i.e.,  
\begin{align}
	\frac{\delta}{\delta y_j(t')}\left[ \dot{y}_i(t)-h_i(x(t))-\mu_i(t) \right]=\delta_{ij}\frac{d}{dt}\delta(t-t').
\end{align}
and can be absorbed into the measure.

Combining these results, it follows that  
\begin{align}
	P(t,x;y|t_0,x_0;y_0)&=\int_{y(t_0)=y_0}^{y(t)=y}\int_{x(t_0)=x_0}^{x(t)=x}[\mc{D}\nu(t)][\mc{D}\mu(t)][\mc{D}x(t)][\mc{D}y(t)]\\ \nonumber 
	&\times\exp\left( -\frac{1}{2\hbar_{\nu}}\sum_{i=1}^n\int_{t_0}^t\nu_i^2(t)dt  -\frac{1}{2\hbar_{\mu}}\sum_{i=1}^n\int_{t_0}^t\mu_i^2(t)dt-\frac{1}{2}\int_{t_0}^t\sum_{i=1}^n\frac{\partial f_i}{\partial x_i}(x(t))dt  \right)\\ \nonumber\
	&\qquad\times \delta^n(\dot{x}(t)-f(x(t))-\nu(t))\delta^m(\dot{y}(t)-h(x(t))-\mu(t)).
\end{align}
The delta functional integrations over $\nu(t)$ and $\mu(t)$ are trivial, so that 
\begin{align}\label{eq:PITimeIndepGen} 
P(t,x;y|t_0,x_0;y_0)=\int_{y(t_0)=y_0}^{y(t)=y}\int_{x(t_0)=x_0}^{x(t)=x}[\mc{D}x(t)][\mc{D}y(t)]\exp(-S),
\end{align}
where 
\begin{align}
S=\frac{1}{2}\int_{t_0}^tdt\left[ \frac{1}{\hbar_{\nu}}\sum_{i=1}^n(\dot{x}_i(t)-f_i(x(t)))^2+\sum_{i=1}^n\frac{\partial f_i}{\partial x_i}(x(t)) +\frac{1}{\hbar_{\mu}}\sum_{k=1}^m(\dot{y}_k-h_k(x(t)))^2  \right].
\end{align}
$S$ is referred to as the action.

\subsection{Sampled Continuous Measurements}\label{ssec:PITISimp}

The path integral formula, Equation \ref{eq:PITimeIndepGen}, can be further simplified by considering the case when the measurements are available at discrete time instants. Specifically, suppose measurements are available at time $t_{i-1}$ and $t_{i}$, and that the transition probability density at time $t_i$ is desired. Further, assume that there are no measurements available between $t_{i-1}$ and $t_i$.  Then, 
\begin{align}
	-\frac{1}{2\hbar_{\mu}}\int_{t_{i-1}}^{t_i}dt\sum_{k=1}^m(\dot{y}_k-h_k(x(t)))^2&=-\frac{1}{2\hbar_{\mu}}\int_{t_{i-1}}^{t_i}dt\sum_{k=1}^m\left[\dot{y}_k^2(t)+h_k^2(x(t))-2h_k(x(t))\dot{y}_i(t)  \right].
\end{align}
The quantity of interest is the state and so all state-independent contributions may be ignored. 

As the first term is independent of the state variables, it is can be absorbed in the measure. The second term can be added to the action term that is independent of $y(t)$. It remains to evaluate the third term:
\begin{align}
		\frac{1}{\hbar_{\mu}}\int_{t_{i-1}}^{t_i} \sum_{k=1}^mh_k(x(t))\dot{y}_k(t)dt.
\end{align}
There are two issues in this evaluation. Firstly, this can be evaluated via the usual integration by parts, but it is important to note that it is valid only for symmetric discretization. Secondly, since the measurements are sampled, there are two possible interpretations:
\begin{align}
		\frac{1}{\hbar_{\mu}}\int_{t_{i-1}}^{t_i} \sum_{k=1}^mh_k(x(t))\dot{y}_k(t)dt=
		\begin{cases}
			\frac{1}{\hbar_{\mu}}\int_{t_{i-1}+\epsilon}^{t_i+\epsilon} \sum_{k=1}^mh_k(x(t))\dot{y}_k(t)dt,\\ 
			\frac{1}{\hbar_{\mu}}\int_{t_{i-1}-\epsilon}^{t_i-\epsilon} \sum_{k=1}^mh_k(x(t))\dot{y}_k(t)dt.
		\end{cases}
\end{align}
This leads to two possibilities:
\begin{align}
		\frac{1}{\hbar_{\mu}}\int_{t_{i-1}}^{t_i} \sum_{k=1}^mh_k(x(t))\dot{y}_k(t)dt=
		\begin{cases}
			\frac{1}{\hbar_{\mu}}\sum_{k=1}^mh_k\left(\frac{1}{2}[x(t_i)+x(t_{i-1})]\right)[y_k(t_i)-y_k(t_{i-1})],\\
			\frac{1}{\hbar_{\mu}}\sum_{k=1}^mh_k\left(\frac{1}{2}[x(t_i)+x(t_{i-1})]\right)[y_k(t_{i-1})-y_k(t_{i-2})].
		\end{cases}
\end{align}
Therefore, the path integral formula simplifies to
\begin{align}\label{eq:PITimeIndepSamp}
P(t_i,x_i;y_i|t_{i-1},x_{i-1};y_{i-1})&=\tilde{P}(t_i,x_i|t_{i-1},x_{i-1})\\ \nonumber
&\qquad\times
\begin{cases}
\exp\left(\frac{1}{\hbar_{\mu}}\sum_{k=1}^mh_k\left(\frac{1}{2}[x(t_i)+x(t_{i-1})]\right)[y_k(t_i)-y_k(t_{i-1})]\right),\\
\exp\left(\frac{1}{\hbar_{\mu}}\sum_{k=1}^mh_k\left(\frac{1}{2}[x(t_i)+x(t_{i-1})]\right)[y_k(t_{i-1})-y_k(t_{i-2})]\right),
\end{cases}
\end{align}
where 
\begin{align}\label{eq:PIFormulaYauKernel}
	\tilde{P}(t_i,x_i|t_{i-1},x_{i-1})=\int_{x(t_{i-1})=x_{i-1}}^{x(t_i)=x_i}[\mc{D}x(t)]\exp(-S(t_{i-1},t_i)),
\end{align}
and
\begin{align}
	S(t_{i-1},t_i)= \frac{1}{2}\int_{t_{i-1}}^{t_i}dt\left[\frac{1}{\hbar_{\nu}}\sum_{i=1}^n(\dot{x}_i(t)-f_i(x(t)))^2+\sum_{i=1}^n\frac{\partial f_i}{\partial x_i}(x(t)) +\frac{1}{\hbar_{\mu}}\sum_{k=1}^mh_k^2(x(t))\right]. 
\end{align}
It shall be shown in Section \ref{sec:DerivYauAlg} that these lead to the two forms of the Yau algorithm.

\section{Path Integral Formula: Explicit Time Dependence}\label{sec:PIFormula02}

In this section, the following model is considered:
\begin{align}
		{\left\lbrace\begin{aligned}
		d\rv{x}(t)&=f(\rv{x}(t),t)dt+e(t)d\rv{\nu}(t),\quad x(0)=x_0,\quad Q(t)\in\mathbb{R}^{n\times n},\\
		d\rv{y}(t)&=h(\rv{x}(t),t)dt+n(t)d\rv{\mu}(t), \quad y(0)=0,\quad R(t)\in\mathbb{R}^{m\times m}. \end{aligned}
		\right.}
\end{align}
The discussion is similar to that in Section \ref{sec:PIFormula01}. However, imposing a delta function constraint requires a different method since $e(t)$ and $n(t)$ are not square matrices, and hence are not invertible. 

\subsection{General Result}

As in Section \ref{sec:PIFormula01}, the transition probability density is computed by averaging over the signal and measurement ensembles, i.e.,
\begin{align}
	P(t,x;y|t_0,x_0;y_0)&=\int[d\rho(\nu(t))][d\rho(\mu(t))]\\ \nonumber
	&\qquad\times[d\dot{x}(t)]J\delta^n[\dot{x}(t)-f(x(t),t)-e(t)\nu(t)] \\ \nonumber
	&\qquad\times [d\dot{y}(t)]J_{y}\delta^m[\dot{y}(t)-h(x(t),t)-n(t)\mu(t)]\\ \nonumber
	&\qquad\times\delta^n[x(t)-x]|_{x(t_0)=x_0}\delta^m[y(t)-y]|_{y(t_0)=y_0}.
\end{align}
From the assumptions of the signal and measurement noise processes, it is evident that  
\begin{align}
	[d\rho(\nu)(t)]&=[\mc{D}\nu(t)]\exp\left( -\frac{1}{2}\sum_{a,b=1}^p\int_{t_0}^t\nu_a(t)\left( Q^{-1}(t) \right)_{ab}\nu_b(t)dt \right), \\ 
	[d\rho(\mu)(t)]&=[\mc{D}\mu(t)]\exp\left( -\frac{1}{2}\sum_{c,d=1}^q\int_{t_0}^t\mu_c(t)\left( W^{-1}(t) \right)_{cd}\mu_d(t)dt \right).
\end{align}
The Jacobian $J$ follows from the functional derivative of the Langevin equation:
\begin{align}
	\frac{\delta}{\delta x_j(t')}\left[ \dot{x}_i(t)-f_i(x(t),t)-\sum_{a=1}^pe_{ia}(t)\nu_a(t) \right]&=\left[ \delta_{ij}\frac{d}{dt}-\frac{\partial f_i}{\partial x_j}(x(t),t) \right]\delta(t-t'),\\ \nonumber
	&=-\frac{d}{dt'}\left[ \delta_{ij}\delta(t-t')-\theta(t-t')\frac{\partial f_i}{\partial x_j}(x(t),t) \right].
\end{align}
Hence, 
\begin{align}
	J&=\det\left( -\frac{d}{dt'} \right)\det\left( \delta(t-t')-\theta(t-t')\frac{\partial f_i}{\partial x_j}(x(t),t) \right),\\ \nonumber
	&=\mc{N}\det\left( \delta(t-t')-\theta(t-t')\frac{\partial f_i}{\partial x_j}(x(t),t)) \right),
\end{align}
where $\mc{N}$ is an irrelevant constant, or, 
\begin{align}
	\ln J&=\ln\det\left[ \delta_{ij}\delta(t-t')-\theta(t-t')\frac{\partial f_i}{\partial x_j}(x(t),t) \right],\\ \nonumber
	&=-\frac{1}{2}\int\sum_{i=1}^n\frac{\partial f_i}{\partial x_i}(x(t),t)dt.
\end{align}
The Jacobian $J_y$ is trivial (as $\dot{y}(t)$ is $y-$independent) and can be absorbed into the measure.

Thus, so far, 
\begin{align}
	P(t,x;y|t_0,x_0;y_0)=&\int_{y(t_0)=y_0}^{y(t)=y}\int_{x(t_0)=x_0}^{x(t)=x}\left[ \mc{D}x(t) \right]\left[ \mc{D}\nu(t) \right]\exp\left( -\frac{1}{2}\int\sum_{i,j=1}^n\sum_{a,b=1}^p\nu_a(t)(Q_{ab}(t))^{-1}\nu_b(t)dt \right)\\ \nonumber
	&\quad\times\delta^n\left(\dot{x}_i(t)-f_i(x(t),t)-\sum_{a=1}^pe_{ia}(t)\nu_a(t)\right)\exp\left( -\frac{1}{2}\int\sum_{i=1}^n\frac{\partial f_i}{\partial x_i}(x(t),t)dt \right)\\
	&\quad\times\delta^m\left( \dot{y}_k(t)-h_k(x(t),t)-\sum_{c=1}^qn_{kc}(t)\mu_c(t) \right).
\end{align}
Using the Fourier integral version of the delta function (Equation \ref{eq:FuncDeltaFn}) 
\begin{align}
	P(t,x;y|t_0,x_0;y_0)=&\int_{y(t_0)=y_0}^{y(t)=y}\int_{x(t_0)=x_0}^{x(t)=x}\left[ \mc{D}x(t) \right]\left[ \mc{D}y(t) \right]\left[ \mc{D}\nu(t) \right][\mc{D}\mu(t)]\\ \nonumber
	&\qquad \left[ \mc{D}\lambda(t) \right]\exp\left( i\int\sum_{i=1}^n\lambda_i(t)\left[\dot{x}_i(t)-f_i(x(t),t)-\sum_{a=1}^pe_{ia}(t)\nu_a(t)\right] \right) \\ \nonumber
	&\qquad \left[ \mc{D}\kappa(t) \right]\exp\left( i\int\sum_{k=1}^m\kappa_k(t)[\dot{y}_k(t)-h_k(x(t),t)-\sum_{c=1}^qn_{kc}(t)\mu_c(t)] \right) \\ \nonumber
	&\times\exp\left( -\frac{1}{2}\int\sum_{a,b=1}^p\nu_a(t)(Q^{-1}(t))_{ab}\nu_b(t)dt \right)\times\exp\left( -\frac{1}{2}\int\sum_{i=1}^n\frac{\partial f_i}{\partial x_i}(x(t),t)dt \right).
\end{align}
Integrating over $\nu(t),$  and $\mu(t)$ leads to  
\begin{align}
P(t,x;y|t_0,x_0;y_0)=&\int_{y(t_0)=y_0}^{y(t)=y}\int_{x(t_0)=x_0}^{x(t)=x}\left[ \mc{D}x(t) \right]\left[ \mc{D}\lambda(t) \right] [\mc{D}\lambda][\mc{D}\mc{\kappa}]\exp\left(-\int dt\left[\frac{1}{2}\sum_{i=1}^n\frac{\partial f_i}{\partial x_i}(x(t),t)\right] \right)\\ \nonumber
	&\times\exp\left( -\int dt\left[ \sum_{i,j=1}^n\sum_{a,b=1}^p\lambda_i(t)e_{ia}(t)Q_{ab}(t)e_{bj}^T(t)\lambda_j(t)-i\sum_{i=1}^n\lambda_i(t)\left( \dot{x}_i(t)-f_i(x(t),t) \right)\right]\right)\\ \nonumber
	&\times\exp\left( -\int dt\left[ \sum_{k,l=1}^m\sum_{c,d=1}^q\kappa_k(t)n_{kc}(t)Q_{cd}(t)n_{dl}^T(t)\kappa_l(t)-i\sum_{k=1}^m\kappa_k(t)\left( \dot{y}_i(t)-h_i(x(t),t) \right)\right]\right).
\end{align}
Integrating over $\lambda_i(t)$ and $\kappa_k(t)$, it is clear that 
\begin{align}\label{eq:TransProbTimeDepFinal}
	P(t,x;y|t_0,x_0;y_0)=\int_{y(t_0)=y_0}^{y(t)=y}\int_{x(t_0)=x_0}^{x(t)=x}[\mc{D}x(t)][\mc{D}y(t)]\exp(-S),
\end{align}
where the action is given by
\begin{align}
	S&= \frac{1}{2}\int dt \sum_{i,j=1}^n\sum_{a,b=1}^p(\dot{x}_i(t)-f_i(x(t),t))\left( e_{ia}(t)Q_{ab}(t)e_{bj}^T(t) \right)^{-1}\left( \dot{x}_j(t)-f_j(x(t),t) \right)\\ \nonumber 
	&+\frac{1}{2}\int dt\left[ \sum_{k,l=1}^n\sum_{c,d=1}^p(\dot{y}_k(t)-h_k(x(t),t))\left( n_{kc}(t)W_{cd}(t)n_{dl}^T(t) \right)^{-1}\left( \dot{y}_l(t)-h_l(x(t),t) \right)+\sum_{i=1}^n \frac{\partial f_i}{\partial x_i}(x(t),t)\right].
\end{align}

\subsection{Sampled Continuous Measurements}

As in Section \ref{ssec:PITISimp}, an attempt is made to reduce the path integral to just the state variables. The general path integral formula (Equation \ref{eq:TransProbTimeDepFinal}) cannot be simplified in the sampled measurement case as easily as in the time dependent case unless some additional assumptions are made. 

First, observe that if $e(t)Q(t)e^T(t)$ and $n(t)W(t)n(t)$ are $\hbar_{\nu}I_{n\times n}$ and $\hbar_{\mu}I_{m\times m}$ and $h(x(t))$  is not explicitly time dependent, the resulting path integral is the same as  Equation \ref{eq:PITimeIndepSamp}. This is the case studied in \cite{S.T.YauS.-TYau2000}. 

Secondly, if the conditions above are relaxed to allowing explicit time dependence of the drift term in the state model, i.e.,  $f(x(t),t)$,  and and $C=(n(t)W(t)n^T(t))$, then the path integral formula  becomes
\begin{align}\label{eq:PITimeDepSamp01}
P(t_i,x_i;y_i|t_{i-1},x_{i-1};y_{i-1})&=\tilde{P}(t_i,x_i|t_{i-1},x_{i-1}) \\ \nonumber
&\quad\times
\begin{cases}
	\exp(\left(\frac{1}{\hbar_{\mu}}\sum_{k,l=1}^mh_k\left(\frac{1}{2}[x(t_i)+x(t_{i-1})]\right)\left(C^{-1}\right)_{kl}[y_l(t_i)-y_k(t_{i-1})]\right),\\
	\exp(\left(\frac{1}{\hbar_{\mu}}\sum_{k,l=1}^mh_k\left(\frac{1}{2}[x(t_i)+x(t_{i-1})]\right)\left(C^{-1}\right)_{kl}[y_l(t_{i-1})-y_k(t_{i-2})]\right),
\end{cases}
\end{align}
where
\begin{align}
\label{eq:PITDepSamp01YauFundSol}
	\tilde{P}(t_i,x_i|t_{i-1},x_{i-1})=\int_{x(t_{i-1})=x_{i-1}}^{x(t_i)=x_i}[\mc{D}x(t)]\exp(-S(t_{i-1},t_i)),
\end{align}
and the action is given by
\begin{align}
	S(t_{i-1},t_i)= \frac{1}{2}\int_{t_{i-1}}^{t_i}dt\left[\frac{1}{\hbar_{\nu}}\sum_{i=1}^n(\dot{x}_i(t)-f_i(x(t),t))^2+\sum_{i=1}^n\frac{\partial f_i}{\partial x_i}(x(t),t) +\frac{1}{\hbar_{\mu}}\int_{t_0}^tdt\sum_{k=1}^mh_k^2(x(t)) \right]. 
\end{align}

Finally, consider the case when $e(t)Q(t)e(t)$ is time-independent and invertible, but otherwise arbitrary, and $C=(n(t)W(t)n^T(t))$. Note that $C$ is a constant, symmetric matrix and assumed to be invertible. The path integral formula  is given by Equations \ref{eq:PITimeDepSamp01} and \ref{eq:PITDepSamp01YauFundSol} and with the action $S(t_{i-1},t_i)$ is given by
\begin{align}\label{eq:ActionYauTDep}
	S(t_{i-1},t_i)&= \frac{1}{2}\int_{t_{i-1}}^{t_i} dt \sum_{i,j=1}^n\sum_{a,b=1}^p(\dot{x}_i(t)-f_i(x(t),t))\left( e_{ia}(t)Q_{ab}(t)e_{bj}^T(t) \right)^{-1}\left( \dot{x}_j(t)-f_j(x(t),t) \right)\\ \nonumber
	&+\frac{1}{2}\int dt\left[ \sum_{k,l=1}^nh_k(x(t))\left( C^{-1}\right)_{kl}h_l(x(t))+\sum_{i=1}^n \frac{\partial f_i}{\partial x_i}(x(t),t)\right].
\end{align}

In Section \ref{ssec:DerivePDE}, the partial differential equation satisfied by $\tilde{P}(t,x|t_0,x_0)$ is derived. This is a straightforward generalization of the YYe to the state model with explicit time dependence. However, it is not rigorously shown that it approximates the solution of the DMZ equation for this case.

\section{Partial Differential Equations Satisfied by the Path Integral Formulas}\label{sec:Verification}

\subsection{Time-Independent Case and the YYe}\label{ssec:VerifyYau}
 In this section, it is proved that the path integral formula derived in the time-dependent case satisfies the YYe. Our discussion follows the method used by Feynman to verify the path integral formula he derived for the fundamental solution of the  Schr\"odinger equation  \cite{R.P.FeynmanandA.R.Hibbs1965}.

From the Chapman-Kolmogorov semi-group property (or equivalently properties of the Gaussian measure), it follows that 
\begin{align}
	\tilde{P}(t+\epsilon,x|t_0,x_0)=\int \tilde{P}(t+\epsilon,x|t,x')\tilde{P}(t,x'|t_0,x_0)\{d^nx'\},
\end{align}
where $\tilde{P}(t,x|t_0,x_0)$ is given by Equation \ref{eq:PIFormulaYauKernel}.  

Now according to Equation \ref{eq:PIFormulaYauKernel}, 
\begin{align}\label{eq:CKVerify}
	\tilde{P}&(t+\epsilon,x|t,x')=\\ \nonumber
	&A\exp\left(-\frac{1}{2\hbar_{\nu}\epsilon}\sum_{i=1}^n\left[ x_i-x_i'-\epsilon f_i(\bar{x}) \right]^2-\epsilon\frac{1}{2}\sum_{i=1}^n\frac{\partial f_i}{\partial x_i}(\bar{x})-\epsilon\frac{1}{2\hbar_{\mu}}\sum_{k=1}^mh_k^2(\bar{x})\right),
\end{align}
where $\bar{x}=\frac{1}{2}(x+x')$. The dominant contribution is when the following condition is satisfied:
\begin{align}
	x-x'-\epsilon f(\bar{x})\approx0.
\end{align}
In this region this may be written as
\begin{align}
	x&=x'+\epsilon f(\bar{x})+\eta,\qquad\text{or}\\ \nonumber
	x&=x'+\epsilon f(x)+\eta,
\end{align}
 where the equalities are valid to $O(\epsilon)$. Substituting this into Equation \ref{eq:CKVerify},  leads to (keeping  terms up to $O(\epsilon)$)
\begin{align}
	\tilde{P}(t+\epsilon,x|t_0,x_0)&=A\int_{-\infty}^{\infty}\exp\left(-\frac{1}{2\hbar_{\nu}\epsilon}\sum_{i=1}^n\eta_i^2\right) \\ \nonumber
	& \quad\times\left( 1-\frac{\epsilon}{2}\sum_{i=1}^n\frac{\partial f_i}{\partial x_i}(x) -\frac{\epsilon}{2\hbar_{\mu}}\sum_{k=1}^mh_k^2(x)\right)\tilde{P}(t,x'|t_0,x_0)\left\{ d^nx' \right\},\\ \nonumber
	&=A\int_{-\infty}^{\infty}\exp\left(-\frac{1}{2\hbar_{\nu}\epsilon}\sum_{i,j=1}^n\eta_i^2\right)& \\ \nonumber
& \quad\times\left( 1-\frac{\epsilon}{2}\sum_{i=1}^n\frac{\partial f_i}{\partial x_i}(x) -\frac{\epsilon}{2\hbar_{\mu}}\sum_{k=1}^mh_k^2(x)\right)\tilde{P}(t,x'|t_0,x_0)\left( 1-\frac{\epsilon}{2}\sum_{i=1}^n\frac{\partial f_i}{\partial x_i}(x) \right)\left\{d^n\eta\right\}, \\ \nonumber
&=A\int_{-\infty}^{\infty}\left\{ d^n\eta \right\}\exp\left(-\frac{1}{2\hbar_{\nu}\epsilon}\sum_{i=1}^n\eta_i^2\right)\left( 1-\epsilon\sum_{i=1}^n\frac{\partial f_i}{\partial x_i}(x) -\frac{\epsilon}{2\hbar_{\mu}}\sum_{k=1}^mh_k^2(x)\right)\\ \nonumber
&\qquad \tilde{P}(t,x-\epsilon f(x,t)-\eta|t_0,x_0).
\end{align}
Note that the Jacobian of the the transformation from $x'$ to $\eta$ to $O(\epsilon)$ is included in the second step.

The constant $A$ is fixed by 
\begin{align}
\label{eq:Anormaliz}
A\int_{-\infty}^{\infty}\exp\left( -\frac{1}{2\hbar_{\nu}\epsilon}\sum_{i=1}^n\eta_{i}^2 \right)\left\{ d^n\eta \right\}=1.
\end{align}
Hence, it follows that
\begin{align}
	A\int_{-\infty}^{\infty}\eta_i\eta_j\exp\left( -\frac{1}{2\hbar_{\nu}\epsilon}\sum_{i=1}^n\eta_i^2 \right)\left\{ d^n\eta \right\}=\hbar_{\nu}\epsilon\delta_{ij}.
\end{align}

The left hand side of the equation is
\begin{align}
	\tilde{P}(t,x'|t_0,x_0)+\epsilon\frac{\partial \tilde{P}}{\partial t}(t,x|t_0,x_0).
\end{align}
The second order expansion (in $\eta$) of  $P(t,x'|t_0,x_0)$ 
\begin{align}
	\tilde{P}(t,x'|t_0,x_0)&=\tilde{P}(t,x-\epsilon f(x,t)-\eta|t_0,x_0),\\ \nonumber
	&=\tilde{P}(t,x|t_0,x_0)\left(1-\epsilon\sum_{i=1}^n\frac{\partial f_i}{\partial x_i}(x,t)-\epsilon\frac{1}{2\hbar_{\mu}}\sum_{k=1}^mh_k^2(x)\right)\\ \nonumber
	&\qquad-\sum_{i=1}^n\left( \epsilon f_i(x)+\eta_i \right)\frac{\partial \tilde{P}}{\partial x_i}(t
	,x|t_0,x_0)+\frac{1}{2}\sum_{i,j=1}^n\eta_i\eta_j\frac{\partial^2\tilde{P}}{\partial x_i\partial x_j}(t,x|t_0,x_0).
\end{align}
Only in terms of $O(\epsilon)$ are of interest here.  The only terms nonvanishing are terms linear in $\epsilon$ and  quadratic in $\eta$. Then it is easy to verify that the diffusion part  of the Yau Eqution follows from the term quadratic in $\eta$, so that 
\begin{align}
{\left\lbrace\begin{aligned}
\frac{\partial \tilde{P}}{\partial t}(t,x|t_0,x_0)&=\frac{\hbar_{\nu}}{2}\sum_{i=1}^n\frac{\partial^2\tilde{P}}{\partial x_i^2}(t,x|t_0,x_0)-\sum_{i=1}^n\frac{\partial}{\partial x_i}\left[ f_i(x)\tilde{P}(t,x|t_0,x_0) \right]-\frac{1}{2\hbar_{\mu}}\sum_{k=1}^mh_k^2(x)\tilde{P}(t,x|t_0,x_0)\\ 
	\tilde{P}(t_0,x|t_0,x_0)&=\delta^n(x-x_0).\end{aligned}
\right.}
\end{align}
Hence, $\tilde{P}$ satisfies the YYe with a delta-function initial condition\footnote{It is straightforward to see that $\tilde{P}(t_0,x|t_0,x_0)=\lim_{\epsilon\ra0}\tilde{P}(t+\epsilon,x|t,x_0)=\delta^n(x-x_0)$.}. 
\subsection{Time-Dependent Case}\label{ssec:DerivePDE}

We now derive the PDE satisfied by the path integral formula derived for the time-dependent case. This is the analog of the YYe in the time-dependent case.

From the Chapman-Kolmogorov semi-group property, it follows that 
\begin{align}
	\tilde{P}(t+\epsilon,x|t_0,x_0)=\int \tilde{P}(t+\epsilon,x|t,x')\tilde{P}(t,x'|t_0,x_0)\{d^nx'\}.
\end{align}
According to Equation \ref{eq:PITimeDepSamp01} and \ref{eq:ActionYauTDep}
\begin{align}\label{eq:CKVerify00}
	\tilde{P}&(t+\epsilon,x|t,x')=\\ \nonumber
	&A\exp\Bigg(-\frac{1}{\epsilon2}\sum_{i,j=1}^n\left[ x-x'-\epsilon f_i(\bar{x},t) \right]\left[ e
	_{ia}(t)Q_{ab}(t)e^T_{bj} \right]\left[ x-x'-\epsilon f(\bar{x},t) \right]\\ \nonumber
	&\qquad-\epsilon\frac{1}{2}\sum_{i=1}^n\frac{\partial f_i}{\partial x_i}(\bar{x},t)-\epsilon\frac{1}{2\hbar_{\mu}}\sum_{k,l=1}^mh_k(\bar{x})\left( C^{-1} \right)_{kl}h_l(\bar{x})\Bigg).
\end{align}
The contribution is negligible unless 
\begin{align} 
	x-x'-\epsilon f(\bar{x},t)\approx0.
\end{align}
so that 
\begin{align} 
	x&=x'+\epsilon f(\bar{x},t)+\eta,\qquad\text{or} \\ \nonumber
	x&=x'+\epsilon f(x,t)+\eta.
\end{align}
where the equalities are valid to $O(\epsilon)$. Substituting this into Equation \ref{eq:CKVerify}, 
\begin{align}\label{eq:TDepVerify}
	\tilde{P}(t+\epsilon,x|t_0,x_0)&=A\int_{-\infty}^{\infty}\exp\left(-\frac{1}{2\epsilon}\sum_{i,j=1}^n\sum_{a,b=1}^p\eta_i(e_{ia}(t)Q_{ab}(t)e^T_{bj}(t))^{-1}\eta_j\right) \\ \nonumber
	& \quad\times\left( 1-\frac{\epsilon}{2}\sum_{i=1}^n\frac{\partial f_i}{\partial x_i}(x,t) -\frac{\epsilon}{2\hbar_{\mu}}\sum_{k,l=1}^mh_k(x)\left( C^{-1} \right)_{kl}h_l(x)\right)\tilde{P}(t,x'|t_0,x_0)\left\{ d^nx \right\},\\ \nonumber
	&=A\int_{-\infty}^{\infty}\exp\left(-\frac{1}{2\epsilon}\sum_{i,j=1}^n\sum_{a,b=1}^p\eta_i(e_{ia}(t)Q_{ab}(t)e^T_{bj}(t))^{-1}\eta_j\right)& \\ \nonumber
	& \quad\times\left( 1-\frac{\epsilon}{2}\sum_{i=1}^n\frac{\partial f_i}{\partial x_i}(x,t) -\frac{\epsilon}{2\hbar_{\mu}}\sum_{k,l=1}^mh_k(x)\left( C^{-1} \right)_{kl}h_l(x)\right)\\ \nonumber
&\qquad\times \tilde{P}(t,x'|t_0,x_0)\left( 1-\frac{\epsilon}{2}\sum_{i=1}^n\frac{\partial f_i}{\partial x_i}(x,t) \right)\left\{d^n\eta\right\}, \\ \nonumber
&=A\int_{-\infty}^{\infty}\exp\left(-\frac{1}{2\epsilon}\sum_{i,j=1}^n\sum_{a,b=1}^p\eta_i(e_{ia}(t)Q_{ab}(t)e^T_{bj}(t))^{-1}\eta_j\right)& \\ \nonumber
& \quad\times\left( 1-\epsilon\sum_{i=1}^n\frac{\partial f_i}{\partial x_i}(x,t) -\frac{\epsilon}{2\hbar_{\mu}}\sum_{k,l=1}^mh_k(x)\left( C^{-1} \right)_{kl}h_l(x)\right)\tilde{P}(t,x-\epsilon f(x,t)-\eta) ,
\end{align}
where the constant $A$ is fixed by 
\begin{align}
	A\int_{-\infty}^{\infty}\left\{ d^{n}\eta \right\}\exp\left( -\frac{1}{2\epsilon}\sum_{i,j}^n\sum_{a,b=1}^p\eta_i(e_{ia}(t)Q_{ab}(t)e_{bj}^T(t))^{-1}\eta_j \right)=1,
\end{align}
so that
\begin{align}
	A\int_{-\infty}^{\infty}\left\{ d^{n}\eta \right\}\eta_k\eta_l\exp\left( -\frac{1}{2\epsilon}\sum_{i,j}^n\sum_{a,b=1}^p\eta_i(e_{ia}(t)Q_{ab}(t)e_{bj}^T(t))^{-1}\eta_j \right)=\epsilon\sum_{a,b=1}^p\left( e_{ka}(t)Q_{ab}(t)e^T_{bl} \right).
\end{align}
The Jacobian of the the transformation from $x'$ to $\eta$ to $O(\epsilon)$ is included in the second step in Equation \ref{eq:TDepVerify}.

The left hand side of the equation is
\begin{align}
	\tilde{P}(t,x'|t_0,x_0)+\epsilon\frac{\partial \tilde{P}}{\partial t}(t,x|t_0,x_0).
\end{align}
To second order in $\eta$,   $\tilde{P}(t,x'|t_0,x_0)$ is 
\begin{align}
	\tilde{P}(t,x'|t_0,x_0)&=\tilde{P}(t,x-\epsilon f(x,t)-\eta|t_0,x_0),\\ \nonumber
	&=\tilde{P}(t,x|t_0,x_0)\left( 1-\epsilon\sum_{i=1}^n\frac{\partial f_i}{\partial x_i}(x,t)-\frac{\epsilon}{2\hbar_{\mu}}\sum_{k,l}^mh_k(x)\left( C^{-1} \right)_{kl}h_l(x)  \right)\\ \nonumber
	&\qquad-\sum_{i=1}^n\left( \epsilon f_i(x,t)+\eta_i \right)\frac{\partial \tilde{P}}{\partial x_i}(t
	,x|t_0,x_0)+\frac{1}{2}\sum_{i,j=1}^n\eta_i\eta_j\frac{\partial^2\tilde{P}}{\partial x_i\partial x_j}(t,x|t_0,x_0).
\end{align}
Since only terms of $O(\epsilon)$ are of interest, the relevant terms are linear in $\epsilon$ and quadratic in $\eta$.Thus, the PDE for the transition probability density is
\begin{align}
	  {\left\lbrace\begin{aligned}
	\frac{\partial \tilde{P}}{\partial t}(t,x|t_0,x_0)=&\frac{1}{2}\sum_{i,j=1}^n\sum_{a,b=1}^p\left( e_{ia}(t)Q_{ab}(t)e^T_{bj}(t) \right)\frac{\partial^2 \tilde{P}}{\partial x_i\partial x_j}(t,x|t_0,x_0)\\ 
	&-\sum_{i=1}^n\frac{\partial}{\partial x_i}\left[ f_i(x,t)\tilde{P}(t,x|t_0,x_0) \right]\\ 
	&-\frac{1}{2\hbar_{\mu}}\sum_{k,l=1}^mh_k(x)\left( C^{-1} \right)_{kl}h_l(x(t))\tilde{P}(t,x|t_0,x_0),\\
	\tilde{P}(t_0,x|t_0,x_0)&=\delta^n(x-x_0).\end{aligned}
\right.}
\end{align}
Clearly, this reduces to the YYe when $e(t)Q(t)e^T(t)=I, \hbar_{\mu}=\hbar_{\nu}=1$ and the drift in the signal model is not explicitly time dependent.

\section{Derivation of the Yau Algorithm}\label{sec:DerivYauAlg}

In Section \ref{sec:PrelimTrProbDen} it was noted that if $v_{i-1}(t_{i-1},x_{i-1})$ is the conditional probability density at time $t_{i-1}$, then the conditional probability density at time $t_i$ is given by
\begin{align}
	v_i(t_i,x_i)=\int P(t_i,x_i;y_i|t_{i-1},x_{i-1};y_{i-1})v_{i-1}(t_{i-1},x_{i-1})\left\{ d^nx_{i-1}\right\}.
\end{align}

First consider the case discussed in Section \ref{sec:PIFormula01}. Then
\begin{align}\label{eq:VerYauAlg01}
	&v_i(t_i,x_i)=\\ \nonumber
	&\begin{cases}
		\int\tilde{P}(t_i,x_i|t_{i-1},x_{i-1})\exp\left( -\frac{1}{\hbar_{\nu}}\sum_{k=1}^mh_k\left(\frac{1}{2}[x(t_{i})+x(t_{i-1})]\right)[y_k(t_i)-y_k(t_{i-1})] \right)v_{i-1}(t_{i-1},x_{i-1})\left\{ d^nx_{i-1} \right\}\\
		\int\tilde{P}(t_i,x_i|t_{i-1},x_{i-1})\exp\left( -\frac{1}{\hbar_{\nu}}\sum_{k=1}^mh_k\left(\frac{1}{2}[x(t_{i})+x(t_{i-1})]\right)[y_k(t_{i-1})-y_k(t_{i-2})] \right)v(t_{i-1},x_{i-1})\left\{ d^nx_{i-1} \right\}.	
	\end{cases}
\end{align}
When $t_i-t_{i-1}$ is small
\begin{align}
	h_k\left(\frac{1}{2}[x(t_{i})+x(t_{i-1})]\right)\approx h_k(x(t_{i-1}))+O(\sqrt{\epsilon}),
\end{align}
and Equation \ref{eq:VerYauAlg01} becomes
\begin{align}
	v_i(t_i,x_i)=
	\begin{cases}
		\int\tilde{P}(t_i,x_i|t_{i-1},x_{i-1})\exp\left( -\frac{1}{\hbar_{\nu}}\sum_{k=1}^mh_k(x(t_{i-1}))[y_k(t_i)-y_k(t_{i-1})] \right)v_{i-1}(t_{i-1},x_{i-1})\left\{ d^nx_{i-1} \right\}\\
		\int\tilde{P}(t_i,x_i|t_{i-1},x_{i-1})\exp\left( -\frac{1}{\hbar_{\nu}}\sum_{k=1}^mh_k(x(t_{i-1}))[y_k(t_{i-1})-y_k(t_{i-2})] \right)v_{i-1}(t_{i-1},x_{i-1})\left\{ d^nx_{i-1} \right\}.	
	\end{cases}
\end{align}
In Section \ref{sec:Verification}, it was  already proved that $\tilde{P}(t_i,x_i|t_{i-1},x_{i-1})$ is the fundamental solution of the YYe. This implies that $v_i(t_i,x)$ is the solution at $t_i$ of
\begin{align} 
{\left\lbrace\begin{aligned}   
\frac{\partial v_i}{\partial t}(t,x)&=\frac{1}{2}\sum_{i=1}^n\frac{\partial^2v_i}{\partial x_i^2}(t,x)-\sum_{i=1}^nf_i(x)\frac{\partial v_i}{\partial x_i}(t,x)-\left( \sum_{i=1}^n\frac{\partial f_i}{\partial x_i}(x)+\frac{1}{2}\sum_{i=1}^mh_i^2(x) \right)v_i(t,x),\\	
v_i(t_{i-1},x)&=v_{i-1}(t_{i-1},x)
\begin{cases}
	\exp\left( \sum_{j=1}^mh_j(x)\left[ y_j(t_i)-y_j(t_{i-1}) \right] \right)\tilde{u}_{1}(\tau_{1},x),\\
	\exp\left( \sum_{j=1}^mh_j(x)\left[ y_j(t_{i-1})-y_j(t_{i-2}) \right] \right)\tilde{u}_{1}(\tau_{1},x),\\
\end{cases}
\end{aligned}
\right.}	
\end{align}
This is precisely the Yau algorithm.  

Likewise, it is straightforward to see that for the general case studied in Section \ref{sec:PIFormula02}, the Yau algorithm is extended to this case as follows: $v_i(t_i,x_i)$ is the solution at $t_i$ of the PDE
\begin{align}
	{\left\lbrace\begin{aligned}   
\frac{\partial v_i}{\partial t}(t,x)&=
\frac{1}{2}\sum_{i,j=1}^n\sum_{a,b=1}^p\left( e_{ia}(t)Q_{ab}(t)e^T_{bj}(t) \right)\frac{\partial^2 v_i}{\partial x_i\partial x_j}(t,x)\\ 
	&-\sum_{i=1}^n\frac{\partial}{\partial x_i}\left[ f_i(x,t)v_i(t,x) \right]\\ 
	&-\frac{1}{2\hbar_{\mu}}\sum_{k,l=1}^mh_k(x)\left( C^{-1} \right)_{kl}h_l(x(t))v_i(t,x),\\
	v_i(t_{i-1},x)&=
	\begin{cases}
		\exp\left( \sum_{j,k=1}^mh_j(x)\left( C^{-1} \right)_{jk}\left[ y_k(t_i)-y_k(t_{i-1}) \right] \right)v_{i-1}(t_{i-1},x),\\
		\exp\left( \sum_{j,k=1}^mh_j(x)\left( C^{-1} \right)_{jk}\left[ y_k(t_{i-1})-y_k(t_{i-2}) \right] \right)v_{i-1}(t_{i-1},x).	
	\end{cases}
	\end{aligned}
\right.}	
\end{align}

\section{Some Remarks}
Following are come remarks on the Feynman path integral solution of the filtering problem:
\begin{itemize}
	\item Note that the Feynman path integral formulation has given a complete and self-contained solution of the continuous-continuous filtering problem. For instance, the DMZ equation  or its variants were not used as an input. On the contrary, the Feynman path integral formula naturally led us to the YYe and the Yau algorithm. 
	\item Unlike the YYe, the PI formula is valid even for time-dependent case. Thus, in principle, one can compute the conditional transition probablity density using the conventional methods (see, for instance, \cite{G.PeterLepage2000}).
	\item The YYe can be viewed as a local expression of the path integral formula. That is, a path integral is a global object, while the PDE is a local one. 
	\item 	In this paper, the signal noise  and measurement noise are assumed to be additive. This is a stronger condition than the orthogonalilty of the diffusion vielbein assumed in \cite{S.T.YauS.-TYau2000}. In a subsequent paper, this case will also be covered. 
	\item It is clear that the robust DMZ equation can also be solved using the transition probability density. However, the quantity of interest, the conditional probability density $\sigma(t,x)$ is computed directly by the path integral formula.  
	\item It is noted that other algorithms can also be solved using the Feynman path integral formulas with obvious changes. For example, the splitting-up approach requires the solution of the FPKfe, which corresponds to the case $h(x)=0$. Note that the FPKfe arises naturally in the solution of the continuous-discrete filtering problem\cite{PAPER1}. Of course, as noted previously, it would be unnecessary since the Yau algorithm has the best numerical properties. What is interesting to note is that the path integral formula naturally leads to the algorithm with the best numerical properties. 
\end{itemize}

\section{Examples}

\subsection{The Dirac-Feynman Approximation}
From the discussion in the previous sections, it is evident that computing the path integral requires computing the Lagrangian, $L$, defined by $S=\int dtL(x,\dot{x},t)$. When the time step is infinitesimal, it is given by Equation 90. Therefore, the simplest and crudest approximation is to use this for non-infinitesimal time step.  Specifically, the fundamental solution of the YYe is approximated as follows:
\begin{align}\label{eq:PIApproxCrude}
	&\tilde{P}(t'',x''|t',x')=\\ \nonumber
	&\exp\left(-\frac{(t''-t')}{2}\left\{\frac{1}{\hbar_{\nu}}\sum_{i=1}^n\left[ \frac{x_i''-x_i'}{t''-t'}-f_i\left( \frac{x''+x'}{2} \right) \right]^2+\sum_{i=1}^n\frac{\partial f_i}{\partial x_i}\left( \frac{x''+x'}{2} \right)+\frac{1}{\hbar_{\mu}}\sum_{k=1}^mh_k^2\left( \frac{x''+x'}{2} \right)\right\}\right).
\end{align}

\subsection{Example 1}
As an example, consider the following continuous-continuous filtering model that has been studied in \cite{C-P.A.Fung1995} (and \cite{LototskyR.MikuleviciusB.L.Rozovskii1997})
\begin{align}\label{eq:ExCCModelCF2}
	d\rv{x}(t)&=a\cos(b\rv{x}(t))dt+\sigma_x d\rv{v}(t),\\ \nonumber
	d\rv{y}(t)&=\tan^{-1}\rv{x}(t)+\sigma_yd\rv{w}(t). 
\end{align}
The Lagrangian for this model is given by
\begin{align}
	\frac{1}{2\sigma_x^2}\left( \dot{x}(t)-a\cos(bx(t)) \right)^2-\frac{ab}{2}\sin(bx(t))+\frac{1}{2\sigma_y^2}\left( \tan^{-1}x(t) \right)^2.
\end{align}
The parameters chosen were as in the reference. Specifically, $a=1.2, b=3, \sigma_x=0.3$, spatial grid spacing $\Delta x=0.01$ and extent $[-1.5,1.5]$, temporal grid spacing $\Delta t=0.01$ with $200$ time steps. 

\FIGURE{
  \centering
  \scalebox{0.75}{ \includegraphics{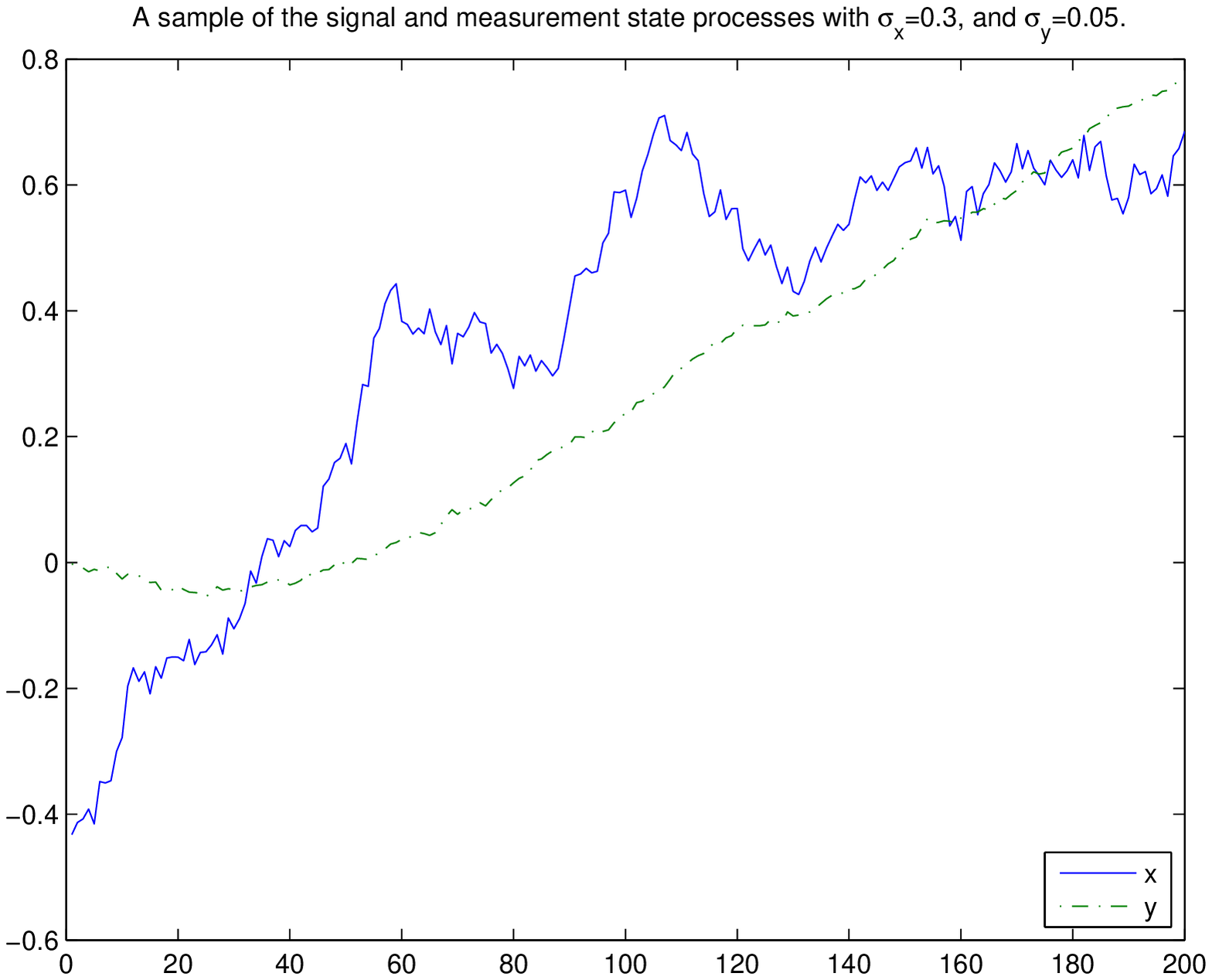}}
 	 \caption{ A sample of state  and measurement processes given by Equation \ref{eq:ExCCModelCF2}.}
        \label{fig:StateMeas5Ex2}
}

In the first set, the measurement noise was set as $\sigma_y=0.05$. Figure \ref{fig:StateMeas5Ex2} shows a sample of state and measurement processes. In Figure \ref{fig:CondMean5} is plotted the conditional mean computed using Equation \ref{eq:PIApproxCrude} in the Yau algorithm\footnote{Since there was negligible difference in performance between the pre-measurement and post-measurement forms, only the former was employed.}. Also plotted are $2\sigma$ limits. The conditional mean and variance were computed from the computed conditional probability density. The fact that the target was mostly within the $2\sigma$ limits of the conditional mean shows that the tracking performance of this algorithm is quite good.

\FIGURE{
  \centering
  \scalebox{0.75}{ \includegraphics{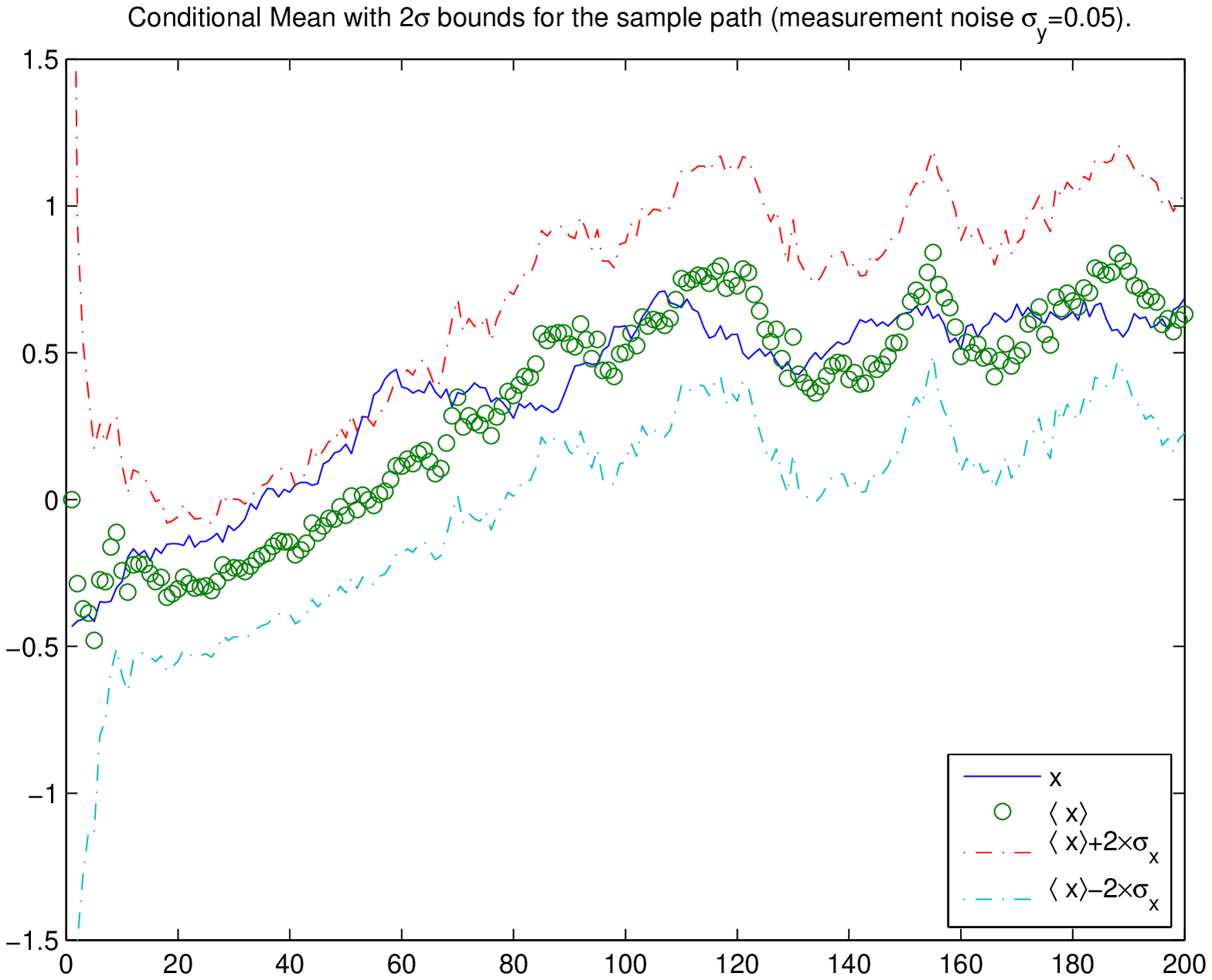}}
 	\caption{Conditional mean and $2\sigma$ limits computed using the Yau algorithm with path integral approximation of Equation \ref{eq:PIApproxCrude}. }
	\label{fig:CondMean5}
}

In the next set, the measurement noise was set as $\sigma_y=0.0125$. For this ``small noise'' case, most of the algorithms studied in \cite{C-P.A.Fung1995} failed. In Figure \ref{fig:StateMeas125Ex2} is plotted a sample of signal and measurement processes. The conditional mean and the $2\sigma$ limits computed using the Yau algorithm for this instance is plotted in Figure \ref{fig:CondMean125}.
\FIGURE{
  \centering
  \scalebox{0.75}{ \includegraphics{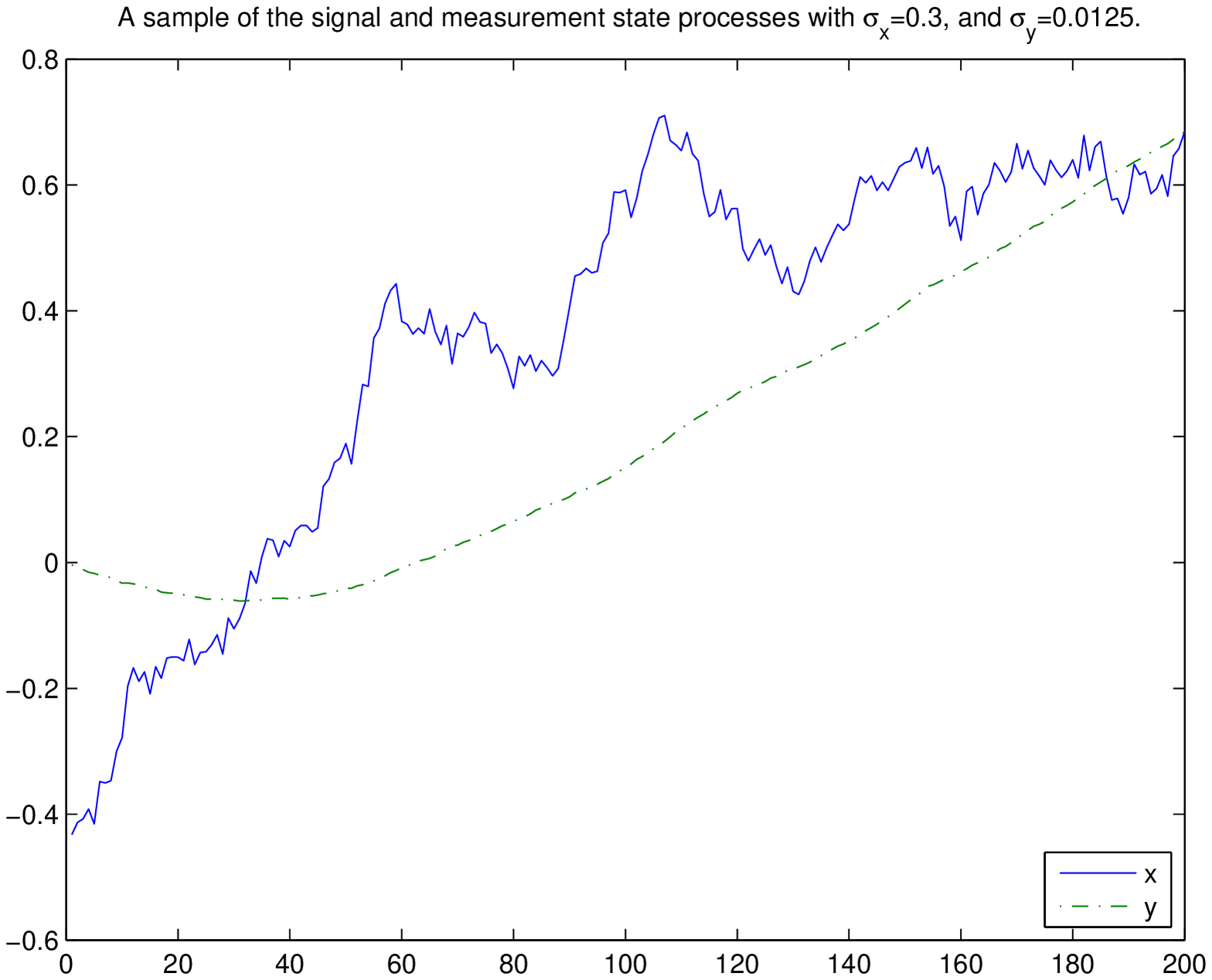}}
 	 \caption{ A sample of state  and measurement processes given by Equation \ref{eq:ExCCModelCF2}---small measurement noise case.}
        \label{fig:StateMeas125Ex2}
}

\FIGURE{
  \centering
  \scalebox{0.75}{ \includegraphics{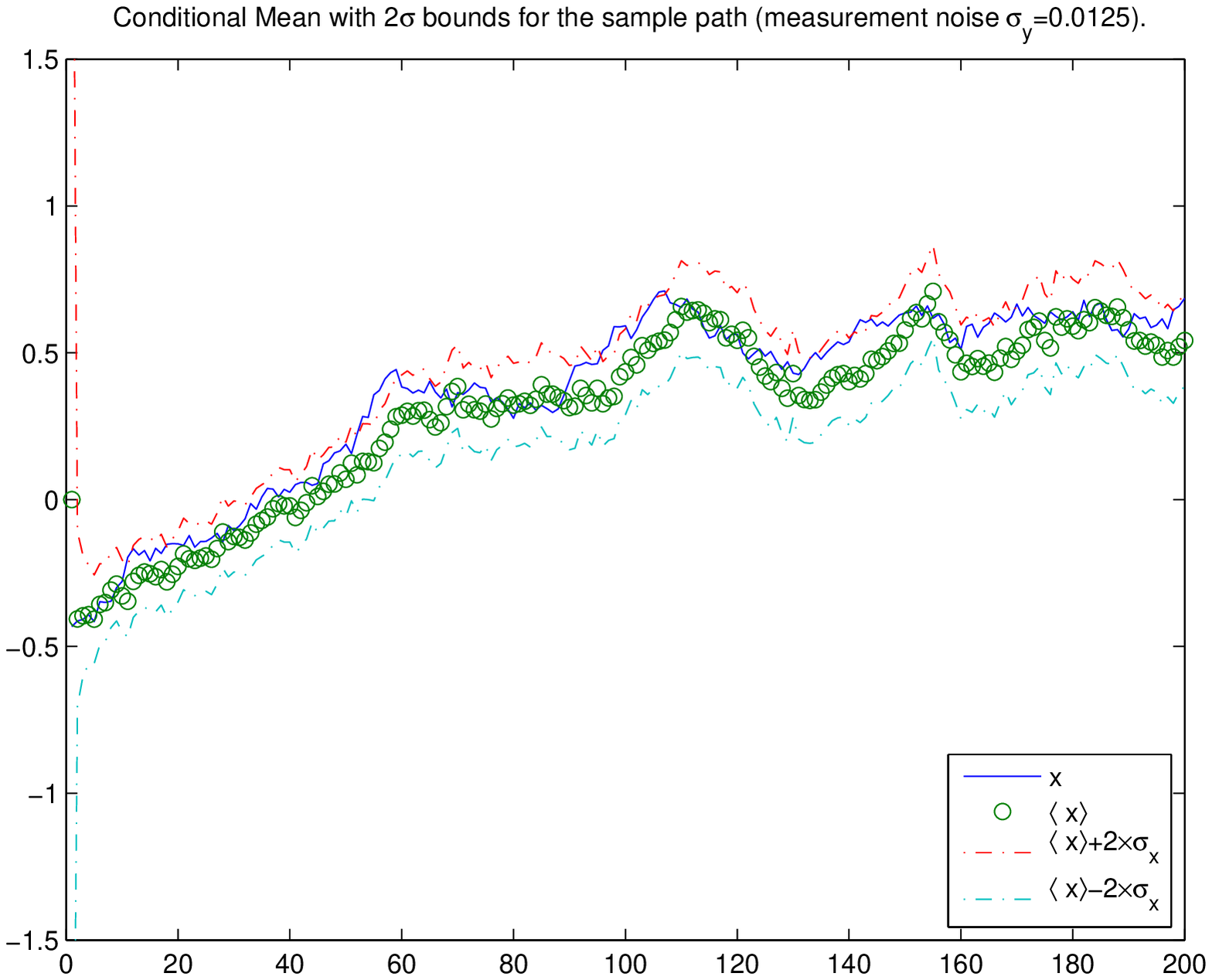}}
 		\caption{Conditional mean and $2\sigma$ limits computed using the Yau algorithm with path integral approximation of Equation \ref{eq:PIApproxCrude} with small measurement noise. }
	\label{fig:CondMean125}
}

It is seen that good tracking performance is maintained for the small noise case even when the crudest path integral approximation is used in the Yau algorithm. 

\subsection{Example 2: Cubic Sensor Problem}
\FIGURE{
  \centering
  \scalebox{0.75}{\includegraphics[scale=0.75]{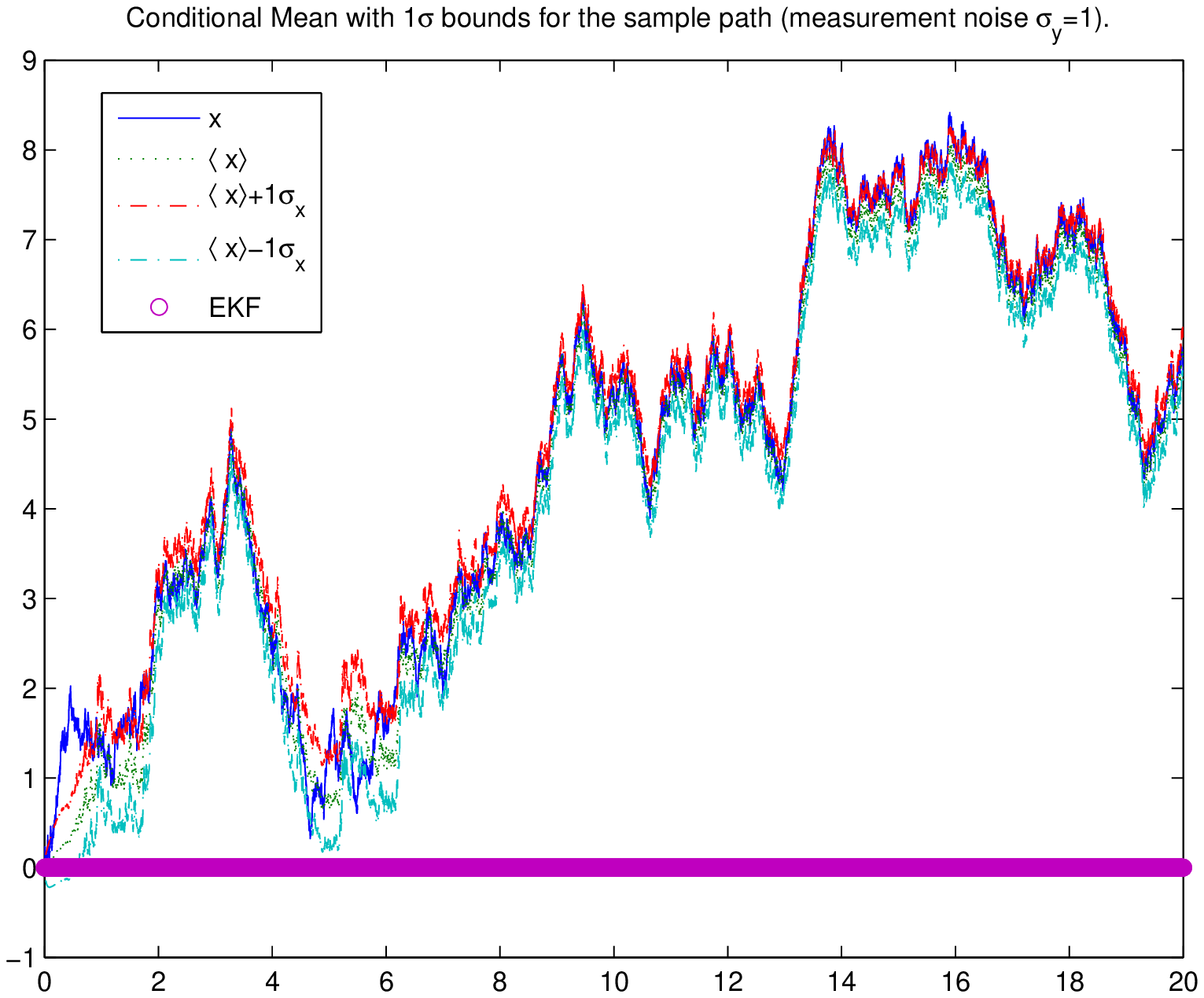}}

	\caption{A simulation result of cubic sensor problem.}
	\label{fig:YauCubicSensor}
	}
The cubic sensor problem is  defined by the following signal and measurement model:
\begin{align}\label{eq:CubicSensorModel}
	d\rv{x}(t)&= \sigma_xd\rv{v}(t),\\ \nonumber
	d\rv{y}(t)&=\rv{x}^3(t)dt+\sigma_yd\rv{w}(t).
\end{align}
It is a well-studied nonlinear filtering problem because it is one of the simplest example examples of a filtering problem that is not finite dimensional (see, for instance, \cite{ChanglinYanandStephenS.-T.Yau2006} and references therein).

For the simulation of the cubic sensor problem, the following model parameters were chosen (as in \cite{ChanglinYanandStephenS.-T.Yau2006}) 
\begin{align}
	\Delta t&=0.001,\\ \nonumber
	\Delta x&=0.01,\\ \nonumber
	x_0&=N(0,0.01),\\ \nonumber
	\sigma_x&=1,\\ \nonumber
	\sigma_y&=1,\\ \nonumber
	T&=[0,20].
\end{align}

The Extended Kalman filter, a sub-optimal filter which approximates the conditional probability by a Gaussian,  for the cubic sensor problem is given by 
\begin{align}
	d\hat{x}&=3\hat{x}^2P(dy-\hat{x}^3dt),\\ \nonumber
	dP&=(1-9\hat{x}^4P^2)dt.
\end{align}

The Lagrangian for the cubic sensor problem is 
\begin{align}
	L=\frac{1}{2\sigma_x^2}(\dot{x}^2)+\frac{1}{2\sigma_y^2}x^6.
\end{align}
The simplest approximation is to use the  simplest approximation of the Lagrangian 
\begin{align}
	\tilde{P}(t'',x''|t',x')=\exp\left( -\frac{\epsilon}{2\sigma_x^2}\left( \frac{x''-x'}{t''-t'} \right)^2-\frac{\epsilon}{2\sigma_y^2} \left( \frac{x''+x'}{2} \right)^6\right).
\end{align}

Figure \ref{fig:YauCubicSensor} shows the performance of the  approximate path integral. Specifically, the conditional mean along with the one standard deviation bounds computed using the computed conditional probability density is plotted.  Observe that the EKF fails completely in this case. As noted in \cite{ChanglinYanandStephenS.-T.Yau2006}, this is because the EKF considers only the first two moments (which vanish here); it is the fourth central moment that plays a crucial role in this example (for the chosen initial condition). Also, note that the state is within the $1\sigma$ region for most of the time. This shows that, unlike the EKF, the path integral filter has a reliable error analysis.  

After an initial period, the performance of the path integral approximation is seen to be excellent and comparable to that obtained using PDE methods in \cite{ChanglinYanandStephenS.-T.Yau2006}. However, the crucial point is that \textit{the path integral solution is equally simple for the higher-dimensional case with more complicated models (e.g., colored noise), whereas a PDE solution would be significantly harder, if not impossible, to implement in real-time.}

\subsection{Comments}

It is remarkable to note the very good performance is obtained using the crudest approximation. Of course, when the time step is large, it will fail (unlike the path integral formula itself). However, the practical situation is that the time steps are often small. Therefore, this may not be a bad approximation in many cases.

The implementation of this method is trivial. The contrast with other methods, such as those studied in  \cite{C-P.A.Fung1995}, is striking. For instance, many of those methods require off-line computation of complicated partial differential equations with uncertain numerical properties.

The results obtained in this paper used single time-step. More accuracy can be obtained quite simply using multiple time steps. Also, the computation of the transition probability density can be done off-line, but the on-line computation was not an onerous burden for this case. 

It is important to note also that the transition probability density matrix (or tensor in the general case) is sparse (determined by $\hbar_{\mu},\hbar_{\nu}$). This  is of great importance in higher dimensional filtering problems because
\begin{itemize}
	\item Sparse matrix storage requirements are small,
	\item The relevant transition probability density matrix elements can be computed based on the conditional  density in the previous step, and
	\item  Sparse matrix computations are very fast.
\end{itemize}

Note that unlike some other approximation techniques studied in \cite{C-P.A.Fung1995}, the conditional probability density is obviously always positive (provided, of course, that the initial distribution is positive).

 \section{Conclusion}\label{sec:Conclusion}

In this paper, the formal path integral solution to the continuous-continuous nonlinear filtering problem has been presented. The solution is universal, i.e., the initial distribution may be arbitrary. We verified that the path integral formula satisfies the YYe when the drift is not explicitly time-independent. Since the path integral measure is manifestly positive, positivity is maintained if the initial distribution is positive.

A path integral formulation has several advantages. Path integrals have led to theoretical insights in other areas including quantum mechanics, quantum field theory and even mathematics. It is demonstrated in this paper that it is quite easy to express the fundamental solution of the YYe in terms of path integrals. In fact, the YYe has been extended to the case when the drift is explicitly time-dependent. In a future paper, this analysis is extended to the case of multiplicative noise\cite{PAPER4}. 

Finally, path integrals are very suitable for numerical implementation. Practical path integral filtering techniques will be presented in subsequent papers. 

\bibliographystyle{utphys}
\bibliography{onfbib}

\providecommand{\href}[2]{#2}\begingroup\raggedright\begin{thebibliography}{10}

\bibitem{A.H.Jazwinski1970}
A.~H. Jazwinski, {\em Stochastic Processes and Filtering Theory}.
\newblock Dover Publications, 2007.

\bibitem{R.E.Kalman1960}
R.~E. Kalman, ``A new approach to linear filtering and prediction problems,''
  {\em Trans. ASME, J. Basic Eng.} {\bf 82D} (March, 1960) 35--45.

\bibitem{R.E.KalmanR.S.Bucy1961}
R.~E. Kalman and R.~S. Bucy, ``New results in linear filtering and prediction
  problems,'' {\em Trans. ASME, J. Basic Eng.} {\bf 83} (1961) 95--108.

\bibitem{H.J.Kushner1964}
H.~J. Kushner, ``On the dynamical equations of conditional probability
  densities of markov processes,'' {\em Siam Journal of Control} {\bf 2} (1964)
  106--119.

\bibitem{H.J.Kushner1964a}
H.~J. Kushner, ``On the differential equations of satisfied by the conditional
  probability densities of markov processes,'' {\em SIAM Journal of Control}
  {\bf 2} (1964) 106--119.

\bibitem{R.L.Stratanovich1960}
R.~L. Stratanovich, ``Conditional markov processes,'' {\em Theor. Prob. Appl.}
  {\bf 5} (1960) 156--178.

\bibitem{R.S.Bucy1965}
R.~S. Bucy, ``Nonlinear filtering theory,'' {\em IEEE Transactions on Automatic
  Control} {\bf 10} (April, 1965) 198.

\bibitem{T.E.Duncan1967}
T.~E. Duncan, {\em Probability densities for diffusion processes with
  applications to nonlinear filtering theory}.
\newblock PhD thesis, Stanford University, 1967.

\bibitem{R.E.Mortensen1966}
R.~E. Mortensen, {\em Optimal control of continuous time stochastic systems}.
\newblock PhD thesis, University of California, Berkeley, 1966.

\bibitem{M.Zakai1969}
M.~Zakai, ``On the optimal filtering of diffusion processes,'' {\em Z. Wahrsch.
  Verw. Geb.} {\bf 11} (1969) 230--243.

\bibitem{M.H.A.Davis1980}
M.~H.~A. Davis, ``On a multiplicative functional transformation arising in
  nonlinear filtering theory,'' {\em Z. Wahrsch Vers. Gebiete} {\bf 54} (1980),
  no.~2, 125--139.

\bibitem{B.L.Rozovskii1972}
B.~L. Rozovskii, ``Stochastic partial differential equations arising in
  nonlinear filtering problems,'' {\em Uspekhi Math. Nauk.} {\bf 27} (1972)
  213--214.

\bibitem{S.T.YauS.-TYau2000}
S.~T. Yau and S.~S.-T. Yau, ``Real time solution of nonlinear filtering problem
  without memory {I},'' {\em Mathematical Research Letters} {\bf 7} (2000)
  671--693.

\bibitem{S.-T.YauS.S.-TYau1996}
S.-T. Yau and S.~S.-T. Yau, ``Explicit solution of a kolmogorov equation,''
  {\em Applied Mathematics and Optimization} {\bf 34} (1996), no.~3, 231--266.

\bibitem{S-T.YauS.S-T.Yau1998}
S.-T. Yau and S.~S.-T. Yau, ``Finite dimensional filters with nonlinear drift
  {XI}: Explicit solution of the generalized kolmogorov equation in
  brockett-mitter program,'' {\em Advances in Mathematics} {\bf 140} (1998)
  156--189.

\bibitem{Balaji2008}
B.~Balaji, ``Estimation of indirectly observable langevin states: path integral
  solution using statistical physics methods,'' {\em Journal of Statistical
  Mechanics: Theory and Experiment} {\bf 2008} (2008), no.~01, P01014 (17pp).

\bibitem{R.P.FeynmanandA.R.Hibbs1965}
R.~P. Feynman and A.~R. Hibbs, {\em Quantum Mechanics and Path Integrals}.
\newblock McGraw-Hill Book Company, 1965.

\bibitem{Feynman1948}
R.~P. Feynman, ``Space-time approach to non-relativistic quantum mechanics,''
  {\em Reviews of Modern Physics} {\bf 20} (1948) 367--387.

\bibitem{P.A.M.Dirac1933}
P.~A.~M. Dirac, ``The lagrangian in quantum mechanics,'' {\em Physikalische
  Zeitschrift der Sowjetunion} {\bf 3} (1933) 64--72.

\bibitem{P.A.M.Dirac1982}
P.~A.~M. Dirac, {\em The Principles of Quantum Mechanics}.
\newblock Oxford University Press, London, UK, fourth~ed., February, 1982.

\bibitem{siegel-1999}
W.~Siegel, ``Fields,'' \href{http://arXiv.org/abs/arXiv:hep-th/9912205v3}{{\tt
  arXiv:hep-th/9912205v3}}.

\bibitem{H.J.Rothe2005}
H.~J. Rothe, {\em Lattice Gauge Theories: An Introduction}.
\newblock World Scientific, third~ed., May, 2005.

\bibitem{Zinn-Justin2002}
J.~Zinn-Justin, ``Quantum field theory and critical phenomena,'' {\em Int. Ser.
  Monogr. Phys.} {\bf 113} (2002)
1--1054.

\bibitem{R.W.BrockettJ.M.C.Clark80}
R.~W. Brockett and J.~M.~C. Clark, {\em The Geometry of the conditional density
  function}, ch.~Analysis and Optimization of Stochastic Systems, pp.~299--309.
\newblock Academic Press, 1980.

\bibitem{R.W.Brockett81}
R.~W. Brockett, {\em Nonlinear systems and nonlinear estimation theory},
  ch.~The Mathematics of Filtering and Identification and Applications,
  pp.~441--477.
\newblock Dordrecht: Reidel, 1981.

\bibitem{Benes1981}
{V.~Ben\v es}, ``Exact finite dimensional filters for certain diffusions with
  nonlinear drift,'' {\em Stochastics} {\bf 5} (1981) 65--92.

\bibitem{S.-T.Yau1994}
S.~S.-T. Yau, ``Finite dimensional filters with nonlinear drift {I}: A class of
  filters including both {K}alman-{B}ucy filters and ben\'es filters,'' {\em J.
  Math. Syst. Estimat. Control} {\bf 4} (April, 1994) 181--203.

\bibitem{YauYau1997}
S.-T. Yau and S.~S.~T. Yau, ``Finite-dimensional filters with nonlinear drift.
  {III}: Duncan-mortensen-zakai equation with arbitrary initial condition for
  the linear filtering system and the benes filtering system,'' {\em IEEE
  Transactions on Aerospace and Electronic Systems} {\bf 33} (Oct., 1997)
  1277--1294.

\bibitem{G.Q.HuS.S-T.Yau2001}
G.~Q. Hu and S.~S.-T. Yau, ``Finite-dimensional filters with nonlinear drift
  {X}: Explicit solution of dmz equation,'' {\em IEEE Transactions on Automatic
  Control} {\bf 46} (2001) 142--148.

\bibitem{G.Q.HuS.S.-T.Yau2002}
G.~Q. Hu and S.~S.-T. Yau, ``Finite dimensional filters with nonlinear drift
  {XV}: New direct method for construction of universal finite-dimensional
  filter,'' {\em IEEE Transactions on Aerospace and Electronic Systems} {\bf
  38} (2002), no.~1, 50--57.

\bibitem{S.-T.YauLai2003}
S.~S.-T. Yau and Y.-T. Lai, ``Explicit solution of {DMZ} equation in nonlinear
  filtering via solution of {ODE}s,'' {\em IEEE Transactions on Automatic
  Control} {\bf 48} (March, 2003) 505--508.

\bibitem{B.Balaji2006}
B.~Balaji, ``Optimal solution of finite dimensional filtering problems via
  solution of linear odes,'' in {\em IEEE Radar Conference}, p.~8.
\newblock 2006.

\bibitem{YauYau2008}
S.-T. Yau and S.~S.-T. Yau, ``Real time solution of the nonlinear filtering
  problem without memory {II},'' {\em SIAM Journal on Control and Optimization}
  {\bf 47} (2008), no.~1, 163--195.

\bibitem{S.S.-T.YauandS.-T.Yau2001}
S.~S.-T. Yau and S.-T. Yau, ``Real time algorithm for nonlinear filtering
  problem,'' in {\em Proceedings of the 40th IEEE Conference on Decision and
  Control}.
\newblock Orlando, FL, USA, December, 2001.

\bibitem{A.BensoussanR.GlowinskiA.Rascanu1990}
A.~Bensoussan, R.~Glowinski, and A.~Rascanu, ``Approximation of the zakai
  equation by the splitting up method,'' {\em SIAM Journal of Control and
  Optiization} {\bf 28} (1990), no.~6, 1420--1431.

\bibitem{R.J.ElliotR.Glowinski1989}
R.~J. Elliot and R.~Glowinski, ``Approximations to solutions of the zakai
  filtering,'' {\em Stochastic Analysis and Applications} {\bf 7} (1989),
  no.~2, 145--168.

\bibitem{P.FlorchingerF.LeGland1991}
P.~Florchinger and F.~LeGland, ``Time discretization of the zakai equation for
  diffusion processes observed in correlated noise,'' {\em Stochastics and
  Stochastics Reports} {\bf 35} (1991), no.~4, 233--256.

\bibitem{LototskyR.MikuleviciusB.L.Rozovskii1997}
S.~Lototsky, R.~Mikulevicius, and B.~L. Rozovskii, ``Nonlinear filtering
  revisited: A spectral approach,'' {\em SIAM Journal of Control and
  Optimization} {\bf 35} (March, 1997) 435--461.

\bibitem{SunGlowinski1993}
M.~Sun and R.~Glowinski, ``Pathwise approximation and simulation for the zakai
  filtering equation through operator splitting,'' {\em Calcolo} {\bf 30}
  (Sept., 1993) 219--239.

\bibitem{PAPER1}
B.~Balaji, ``Universal nonlinear filtering using path integrals {I}: The
  continuous-discrete model with additive noise,'' {\em submitted to IEEE
  Transactions on Aerospace and Electronic Systems} (2006)
  \href{http://arXiv.org/abs/arXiv:0708.0354}{{\tt arXiv:0708.0354}}.

\bibitem{Shing-TungYauStephenS.T.Yau1998}
S.-T. Yau and S.~S.~T. Yau, ``Existence and decay estimates for time dependent
  parabolic equation with application to duncan-mortensen-zakai equation,''
  {\em Asian Journal of Mathematics} {\bf 2} (December, 1998) 1079--1150.

\bibitem{H.Risken1999}
H.~Risken, {\em The {Fokker-Planck} Equation: Methods of Solution and
  applications}.
\newblock Springer-Verlag, second~ed., 1999.

\bibitem{G.PeterLepage2000}
G.~P. Lepage, ``Lattice {QCD} for novices,''
  \href{http://arXiv.org/abs/arXiv:hep-lat/0506036}{{\tt
  arXiv:hep-lat/0506036}}.

\bibitem{C-P.A.Fung1995}
C.-P.~A. Fung, {\em New Numerical algorithms for Nonlinear Filtering}.
\newblock PhD thesis, University of Southern California, 1995.

\bibitem{ChanglinYanandStephenS.-T.Yau2006}
C.~Yan and S.~S.-T. Yau, ``A new suboptimal filter and numerical solutions for
  the cubic sensor problem,'' in {\em Proceedings of the 2006 IEEE
  International Conference on Networking, Sensing and Control}.
\newblock 2006.

\bibitem{PAPER4}
B.~Balaji, ``Universal nonlinear filtering using path integrals {IV}: The
  continuous-continuous model with multiplicative noise,'' {\em in
  preparation}.

\end{thebibliography}\endgroup

\end{document}